\documentclass[12pt,prd,aps,onecolumn,nofootinbib]{revtex4-1}%
\usepackage[colorlinks=true,linkcolor=blue,urlcolor=blue,filecolor=black,citecolor=red,pdfstartview=FitV,pdftitle={},pdfsubject={},
pdfkeywords={},pdfpagemode=None,bookmarksopen=true]{hyperref}
\usepackage{graphicx}%Include figure filespp
\usepackage{epstopdf}%
\usepackage{amsmath}
\usepackage{amsfonts}
\usepackage{amssymb}
\usepackage{xcolor}
\usepackage{color}%
\usepackage{dcolumn}% Align table columns on decimal pointppp
\usepackage{slashed}
\usepackage[normalem]{ulem}
\usepackage{amsthm}
\usepackage{mathrsfs}
\usepackage[caption=false]{subfig}
\usepackage{wrapfig}
\usepackage{indentfirst}

\providecommand{\U}[1]{\protect\rule{.1in}{.1in}}

\begin{document}

\title{Spin-Induced Nonlinear Scalarization of Kerr Black Holes in Einstein-scalar-Gauss-Bonnet Gravity}

\author{Meng-Yun Lai$^{1}$\footnote{ mengyunlai@jxnu.edu.cn}, Hyat Huang$^{1}$\footnote{hyat@mail.bnu.edu.cn}, Jutta Kunz$^{2}$\footnote{jutta.kunz@uni-oldenburg.de},\\ Yun Soo Myung$^{3}$\footnote{ysmyung@inje.ac.kr} and De-Cheng Zou$^{1}$\footnote{dczou@jxnu.edu.cn}}

\address{$^{1}$ School of Physics, Jiangxi Normal University, Nanchang 330022, China \\
$^{2}$Institut f\"ur  Physik, Universit\"at Oldenburg, Postfach 2503, D-26111 Oldenburg, Germany\\
$^{3}$Center for Quantum Spacetime, Sogang University, Seoul 04107, Republic of  Korea}

\begin{abstract}
We investigate spin-induced scalarization of Kerr black holes in an Einstein-scalar-Gauss-Bonnet (EsGB) model that does not admit a linear tachyonic instability of the scalar-free solution. 
The scalarization mechanism is therefore genuinely nonlinear. 
We first analyze the decoupled scalar dynamics on fixed Kerr backgrounds and show that sufficiently rapid rotation modifies the Gauss-Bonnet invariant such that a negative near-horizon region develops near the poles. 
This region provides a geometric trapping mechanism for nonlinear scalar growth, which becomes effective above a threshold spin $\chi=0.5$. 
We then construct stationary scalarized black hole solutions with full backreaction and determine their domain of existence. 
We find that the solutions occupy a finite low-mass high-spin wedge in the spin-mass plane. 
This is in contrast to spin-induced spontaneous scalarization, where the scalarized solutions form a narrow band. 
In this wedge, toward the high-spin end, the scalar hair becomes stronger, and the solutions approach a near-extremal regime, while toward the low-spin boundary, the scalar field is strongly suppressed and approaches a weak-hair limit as $\chi \to 0.5$.

\end{abstract}

\maketitle

\section{Introduction}

Observations of gravitational waves and black hole images provide compelling evidence for the existence of black holes in nature~\cite{LIGOScientific:2016aoc,LIGOScientific:2018jsj,EventHorizonTelescope:2019dse,EventHorizonTelescope:2022wkp}. In general relativity, these objects are expected to obey strong no-hair constraints~\cite{Israel:1967wq,Carter:1971zc,Robinson:1975bv,Bekenstein:1995un,Sotiriou:2015pka,Herdeiro:2015waa}, but these can be circumvented in Einstein-scalar-Gauss-Bonnet (EsGB) theories by a direct coupling between the scalar field and the Gauss-Bonnet invariant~\cite{Kanti:1995vq,Sotiriou:2013qea,Antoniou:2017acq}.
This coupling allows for scalarized black-hole solutions and has led to a broad body of work on spontaneous scalarization in both static and rotating settings~\cite{Doneva:2017bvd,Silva:2017uqg,Cunha:2019dwb,Collodel:2019kkx,Dima:2020yac,Herdeiro:2020wei,Berti:2020kgk}. Similar scalarization mechanisms have also been investigated in related models, including Einstein-Maxwell-scalar theories~\cite{Herdeiro:2018wub,Myung:2018vug,Myung:2018jvi,Fernandes:2019rez,Astefanesei:2019pfq,Blazquez-Salcedo:2020nhs,Chen:2026olq}, Einstein-Chern-Simons-scalar theories~\cite{Doneva:2021dcc,Fan:2023jhi}, and other black hole backgrounds~\cite{Liu:2024bzh,Zhang:2024bfu}. 
In the rotating EsGB case, scalarized Kerr black holes were first constructed with an exponential coupling function~\cite{Cunha:2019dwb}, and later with a simple quadratic coupling~\cite{Collodel:2019kkx}.
Subsequent studies then showed that sufficiently rapid rotation can also trigger spontaneous scalarization, leading to spin-induced scalarized black holes beyond the Kerr family~\cite{Dima:2020yac,Herdeiro:2020wei,Berti:2020kgk}. 
This phenomenon can be traced to the fact that rotation changes the sign structure of the Gauss-Bonnet invariant and creates a favorable polar region for scalar growth~\cite{Hod:2020jjy}.

Scalarization, however, need not always be of the spontaneous type.
In the standard spontaneous scalarization scenario, the scalarized branch is tied to a linear tachyonic instability of the bald black hole background and is therefore associated with the existence of a linear zero mode~\cite{Doneva:2017bvd,Silva:2017uqg}. 
By contrast, in nonlinear scalarization, the bald solution can remain linearly stable, while scalarized black holes may still form under sufficiently large nonlinear scalar perturbations and need not be continuously connected to the bald branch~\cite{Doneva:2021tvn,Doneva:2022yqu,Lai:2023gwe,Zou:2024wsk,Liu:2025eve}. 
This raises the question of whether the spin-induced mechanism can still support scalarization in the absence of a linear tachyonic onset, and, if so, what the corresponding branch structure of rotating scalarized black holes looks like.

In this work, we address this question in an EsGB model with a coupling function $\zeta(\phi)$ satisfying $\zeta''(0)=0$, for which the linearized scalar equation on a Kerr background reduces to the massless wave equation and therefore does not admit a tachyonic instability.
This setup allows us to isolate genuinely nonlinear effects in the scalar sector and to investigate whether rotation alone can trigger scalarization beyond the standard spontaneous scenario~\cite{Dima:2020yac,Herdeiro:2020wei,Berti:2020kgk}. 

Our analysis proceeds in two steps. 
First, we study decoupled scalar dynamics on fixed Kerr backgrounds in order to provide dynamical intuition for the nonlinear growth mechanism.
We show that sufficiently rapid rotation alters the Gauss-Bonnet invariant such that a negative near-horizon region develops near the poles.
This, then, provides a favorable geometric environment for the growth of the scalar field.
In particular, this effect becomes significant only above a threshold spin, $\chi=0.5$.
This can be understood as a geometric condition for the emergence of a polar trapping region.

We then construct stationary scalarized black hole solutions with full back reaction and determine their domain of existence.
In the spin-induced spontaneous scalarization scenario, the domain of existence of scalarized solutions starts at $\chi=0.5$, where the bifurcation line from Kerr black holes arises.
The scalarized solutions then exist in a narrow band in the spin-mass plane, limited toward lower masses by the critical line and extending into the high-spin region by slightly exceeding the Kerr bound.
In contrast, in the nonlinear scalarization case, we find that the solutions occupy a wedge in the spin-mass plane, delimited by vanishing mass and the extremal Kerr spin.

These results provide a global picture of spin-induced nonlinear scalarized black holes and show that rotation can trigger scalarization purely through nonlinear effects without requiring a linear tachyonic instability.
In this sense, the present scenario differs qualitatively from the standard spontaneous scalarization picture and is more closely related to the nonlinear scalarization framework discussed in~\cite{Doneva:2021tvn,Lai:2023gwe}.

\section{Theoretical setup}

\subsection{Einstein-scalar-Gauss-Bonnet model}

We consider Einstein-scalar-Gauss-Bonnet (EsGB) theory with action
\begin{eqnarray}
S = \frac{1}{16\pi} \int d^4x\,\sqrt{-g}
\left[
R - 2\nabla_\mu\phi\nabla^\mu\phi + \alpha \zeta(\phi)\,\mathcal{G}
\right] .
\end{eqnarray}
The last term in the action introduces a direct coupling between the scalar field and the Gauss-Bonnet invariant
\begin{eqnarray}
\mathcal{G} = R_{\mu\nu\rho\sigma}R^{\mu\nu\rho\sigma}
- 4 R_{\mu\nu}R^{\mu\nu} + R^2 ,
\end{eqnarray}
which acts as an effective source for the scalar field.
In particular, the scalar field equation acquires a curvature-induced term proportional to $\zeta'(\phi) \mathcal{G}$, so the sign and magnitude of the Gauss-Bonnet invariant play a crucial role in determining whether the growth of the scalar field is favored.
This mechanism underlies the appearance of scalarized black hole solutions in EsGB theories.
We work in geometrized units $G=c=1$.

Varying the action with respect to the metric and the scalar field yields the coupled set of field equations
\begin{eqnarray}
R_{\mu\nu} - \frac{1}{2} g_{\mu\nu} R &=& \frac{1}{2} T_{\mu\nu} , \\
\Box \phi+ \frac{\alpha}{4} \zeta'(\phi)\,\mathcal{G}  &=& 0,\label{scalarfieldEq0}
\end{eqnarray}
where the effective energy-momentum tensor can be written as
\begin{eqnarray}
T_{\mu\nu} = 4 T^{(\phi)}_{\mu\nu} - \alpha \, T^{(\mathcal{G})}_{\mu\nu} .
\end{eqnarray}
The scalar kinetic contribution is
\begin{eqnarray}
T^{(\phi)}_{\mu\nu}
=
\nabla_\mu \phi \, \nabla_\nu \phi
- \frac{1}{2} g_{\mu\nu}
\nabla_\alpha \phi \, \nabla^\alpha \phi ,
\end{eqnarray}
and the Gauss-Bonnet contribution is
\begin{eqnarray}
T^{(\mathcal{G})}_{\mu\nu}
=
2 g_{\alpha(\mu} g_{\nu)\beta}
\epsilon^{\gamma\beta\delta\kappa}
\nabla_\lambda
\left[
{}^\ast R^{\alpha\lambda}{}_{\delta\kappa}
\, \zeta_{,\phi}(\phi)
\, \nabla_\gamma \phi
\right] ,
\end{eqnarray}
with ${}^\ast R^{\mu\nu}{}_{\rho\sigma}=\epsilon^{\mu\nu\alpha\beta}R_{\alpha\beta\rho\sigma}$ the dual Riemann tensor.

In the present paper, we adopt the coupling function
\begin{eqnarray}
\zeta(\phi)=\frac{1}{4\beta}\left(1-e^{-\beta\phi^4}\right) .
\end{eqnarray}
This choice satisfies $\zeta'(0)=0$ and $\zeta''(0)=0$.
Therefore, the linearized scalar equation around the trivial solution reduces to the massless wave equation on a Kerr background. 
As a result, no tachyonic instability is present at the linear level, and any scalarization must arise from genuinely nonlinear effects.
This makes the model particularly suitable for investigating nonlinear scalarization mechanisms~\cite{Doneva:2021tvn,Lai:2023gwe}. 

Throughout this work, we restrict ourselves to negative Gauss-Bonnet coupling, $\alpha<0$.
In this case, regions where the Gauss-Bonnet invariant becomes negative can provide an effective potential well for the scalar field, thereby favoring the growth of the scalar field.
As will be discussed below, this condition is naturally realized near the poles of rapidly rotating Kerr black holes.
In the numerical calculations, we fix the coupling parameter to $\beta=1000$  and focus on stationary, axisymmetric black hole solutions with nontrivial scalar hair.
We investigate how the interplay between rotation and the curvature coupling leads to scalarization in the absence of a linear instability.
Before constructing solutions with full backreaction, it is instructive to analyze the scalar field dynamics on a fixed Kerr background. 
This decoupling limit provides useful intuition for the nonlinear mechanism responsible for the growth of the scalar field, which we explore in the next section.

\section{Decoupled scalar dynamics on a fixed Kerr background}

On a fixed Kerr background, with the dimensionless spin $\chi\equiv J/M^2$, and outer horizon located at
\begin{eqnarray}
r_+=M(1+\sqrt{1-\chi^2}),
\end{eqnarray}
we study the decoupled evolution of the scalar field, neglecting metric backreaction. 
This approximation allows us to isolate the scalar dynamics and to gain intuition for the nonlinear mechanism responsible for scalar growth in the full system. 
The scalar field equation \eqref{scalarfieldEq0} then reduces to
\begin{eqnarray}
\bar{\Box}\phi+\frac{\alpha}{4}\,\zeta'(\phi)\,\bar{\mathcal G}=0,
\end{eqnarray}
where $\bar{\Box}$ and $\bar{\mathcal G}$ are respectively the d'Alembertian operator and the Gauss-Bonnet invariant evaluated on the Kerr background. 
In this decoupling limit, the scalar field evolves in a fixed curved background, with the Gauss-Bonnet invariant acting as an effective source term.
Since the coupling function satisfies $\zeta'(0)=0$ and $\zeta''(0)=0$, the scalar-free configuration is linearly stable.
Thus any growth of the scalar field must be driven by nonlinear effects.

It is useful to introduce the effective potential
\begin{eqnarray} \label{pot-e}
V_{\rm eff}(\phi,r,\theta)=-\frac{\alpha}{4}\,\zeta(\phi)\,\bar{\mathcal G} ,
\end{eqnarray}
whose derivative with respect to $\phi$ is $\frac{\partial V_{\rm eff}}{\partial\phi}=-\frac{\alpha}{4}\,\zeta'(\phi)\,\bar{\mathcal G}$.
This determines the nonlinear driving term in the scalar field equation.
For $\alpha<0$ and $\zeta'(\phi)>0$, regions where the Gauss-Bonnet invariant satisfies $\bar{\mathcal G}<0$ possess negative values of $V_{\rm eff}$, and therefore provide a favorable environment for the growth of the scalar field.
Thus, the sign structure of $\bar{\mathcal G}$ plays a central role in determining where nonlinear scalarization can occur.

\begin{figure}[t]
\subfloat[$\theta=0$]{\includegraphics[width=0.28\textwidth]{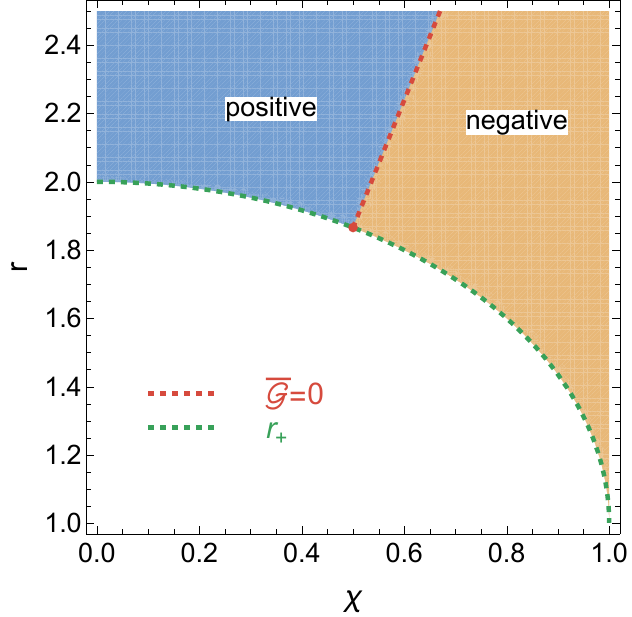}}
\hfill
\subfloat[$\theta=\frac{\pi}{3}$]{\includegraphics[width=0.28\textwidth]{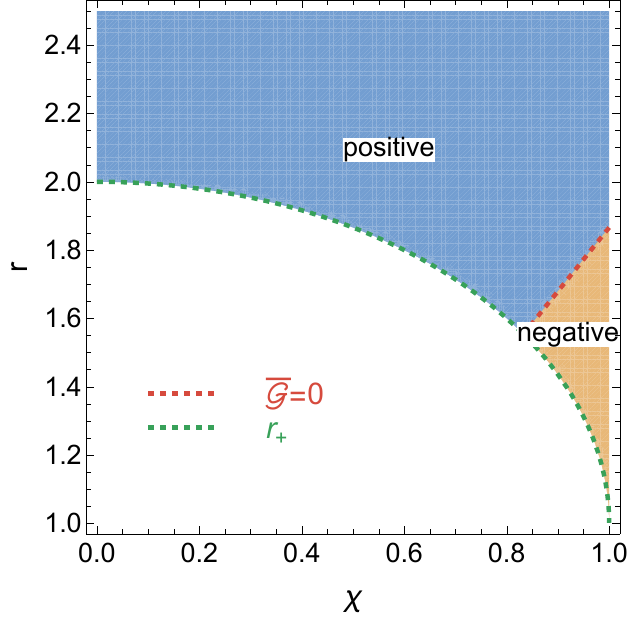}}
\hfill
\subfloat[$\theta=\frac{\pi}{2}$]{\includegraphics[width=0.28\textwidth]{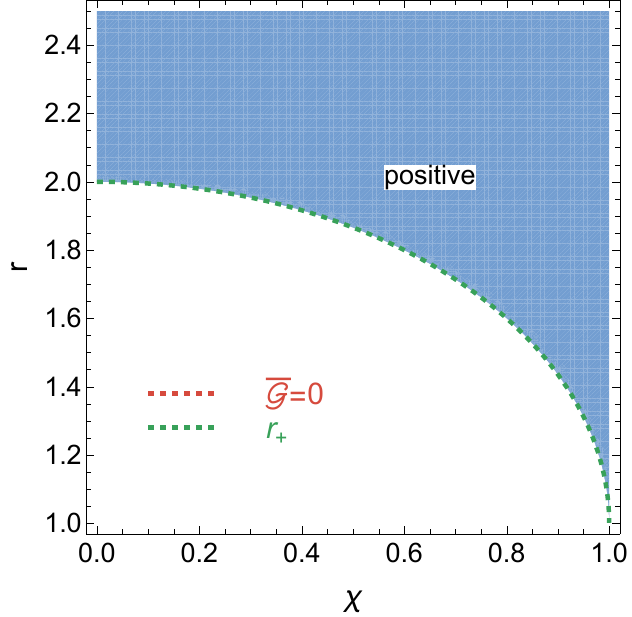}}
\caption{Sign structure of the Kerr Gauss-Bonnet invariant $\bar{\mathcal G}(r,\theta)$ for $M=1$, shown in the $(\chi,r)$ plane for three representative angles: (a) $\theta=0$, (b) $\theta=\pi/3$, and (c) $\theta=\pi/2$. The red dashed line denotes $\bar{\mathcal G}=0$, and the green dashed line denotes the Kerr horizon $r_+$. A negative near-horizon region first appears near the poles when $\chi>0.5$, while no such region is present on the equatorial plane.}
\label{figGsign}
\end{figure}

Figure \ref{figGsign} shows the sign structure of the Gauss-Bonnet invariant $\bar{\mathcal G}(r,\theta)$ for several representative angles. 
A key feature is the emergence of a negative near-horizon region in the polar direction once the spin exceeds a critical value. 
For $\theta = 0$, this region first appears when $\chi  > 0.5$, while for larger polar angles the corresponding threshold increases (e.g., $\chi\simeq0.83$ for $\theta=\pi/3$), and no such region is present on the equatorial plane. 
This behavior indicates that the polar region plays a dominant role in the scalarization process. 
In particular, the appearance of a negative near-horizon region can be interpreted as the formation of a geometric trapping zone for the scalar field.
This suggests that $\chi=0.5$ should be understood as a geometric threshold for nonlinear growth of the scalar field, rather than as a linear instability onset. 
Still, in both cases spin-induced scalarization exhibits the same threshold value.

\begin{figure}[t]
\centering
\subfloat[Effective potential for $\chi=0.3$\label{figVeff_a0d3}]{
    \includegraphics[width=0.45\textwidth]{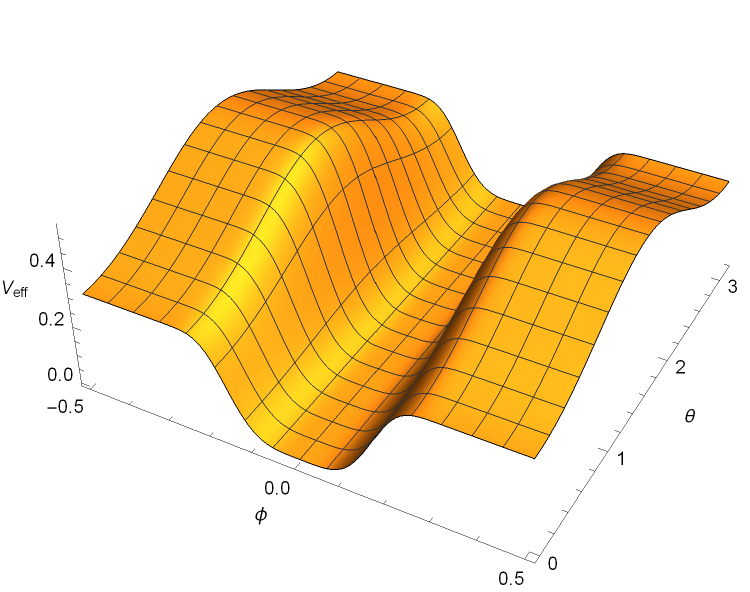}
}
\hfill
\subfloat[Effective potential for $\chi=0.7$\label{figVeff_a0d7}]{
    \includegraphics[width=0.45\textwidth]{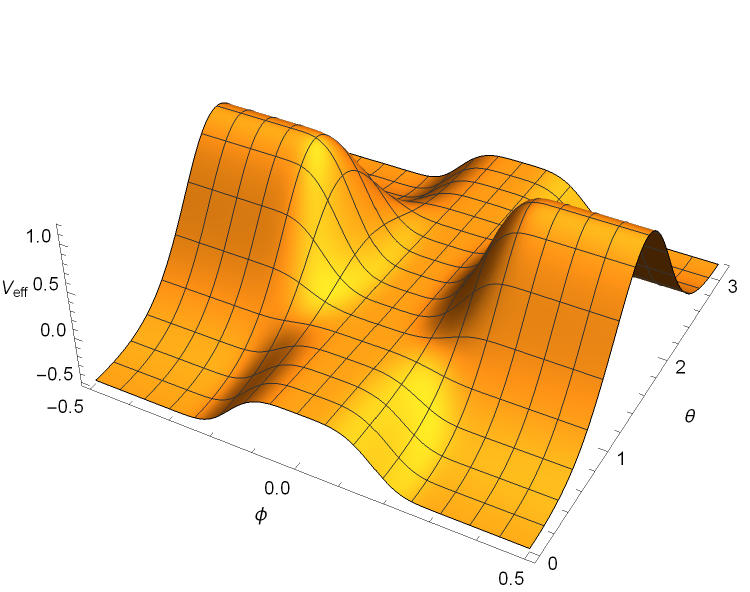}
}

\vspace{0.5cm}

\subfloat[Decoupled scalar evolution for $\chi=0.3$\label{fig_tphi_a0d3}]{
    \includegraphics[width=0.45\textwidth]{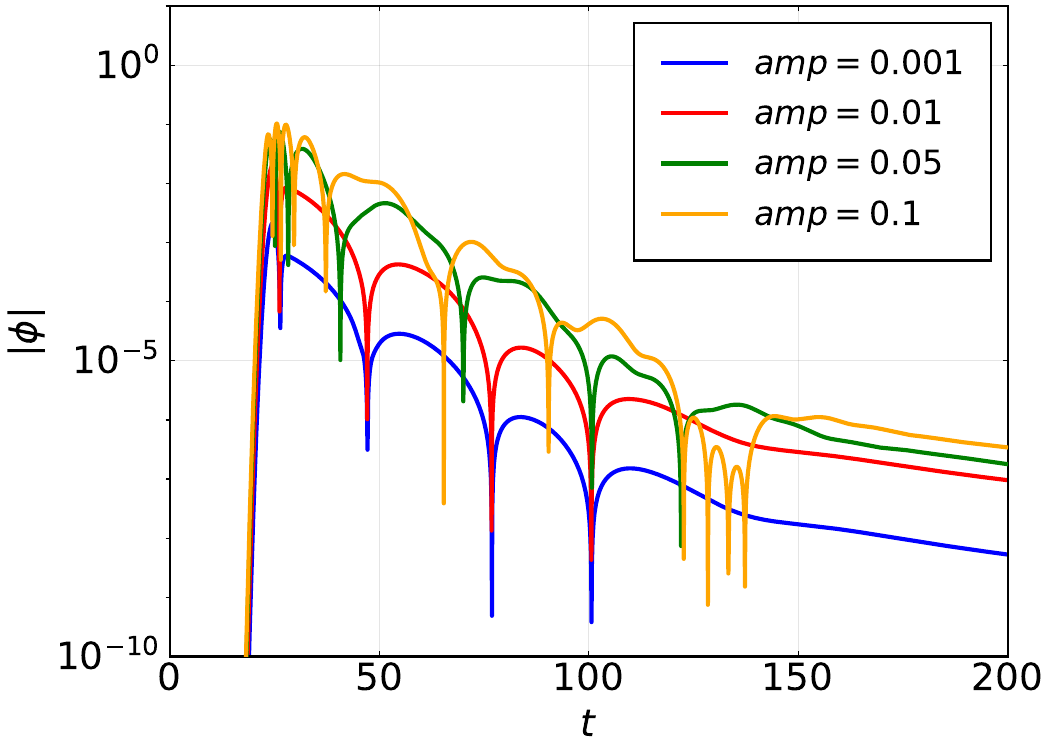}
}
\hfill
\subfloat[Decoupled scalar evolution for $\chi=0.7$\label{fig_tphi_a0d7}]{
    \includegraphics[width=0.45\textwidth]{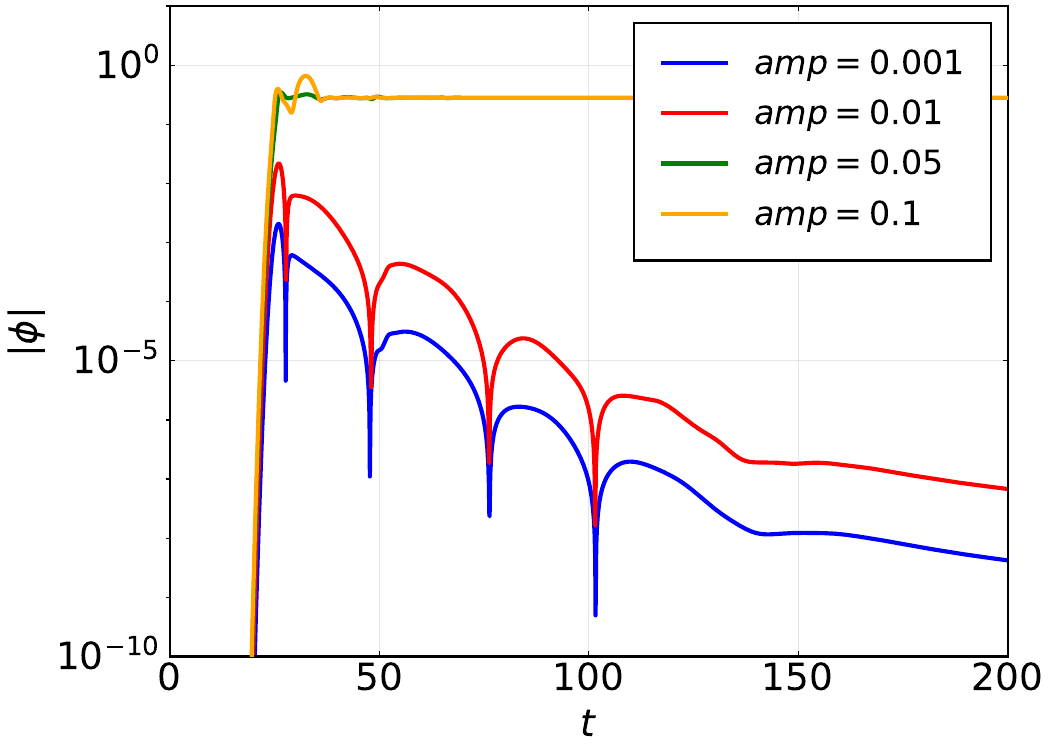}
}

\caption{
Representative decoupled scalar dynamics on fixed Kerr backgrounds, shown here for $\beta=1000$ and $\hat M=0.01$. Panels (a) and (b) display the effective potential $V_{\rm eff}$ for $\chi=0.3$ and $\chi=0.7$, respectively. Panels (c) and (d) show the corresponding scalar time evolutions for several values of $\mathrm{amp}$, where $\mathrm{amp}$ denotes the amplitude of the initial scalar wave packet. For $\chi=0.3$, all perturbations decay back to the scalar-free state. For $\chi=0.7$, sufficiently large amplitudes trigger nonlinear scalar condensation.
}
\label{fig_Veff_tphi}
\end{figure}

To further explore this mechanism, we solve the scalar field equation on the fixed Kerr background and study its time evolution in the decoupling limit. 
The initial scalar perturbation is taken to be a localized wave packet with amplitude denoted by $amp$, which controls the strength of the nonlinear effects. 
Since the system does not exhibit a linear instability, the amplitude of the initial perturbation plays a crucial role in determining whether growth of the scalar field is triggered. 
For the examples shown in Fig.~\ref{fig_Veff_tphi}, we choose parameters corresponding to a small dimensionless mass, $\hat M=M/\sqrt{|\alpha|}=0.01$.

The decoupled dynamics is largely governed by the structure of the effective potential $V_{\rm eff}$.
As the spin increases, the Gauss-Bonnet invariant develops a more pronounced negative region near the poles, making $V_{\rm eff}$ increasingly favorable for the growth of the scalar field. 
As a result, high-spin Kerr backgrounds can support nonlinear scalar condensation, while low-spin backgrounds tend to drive the scalar field back to the trivial configuration.

This behavior is illustrated in Fig.~\ref{fig_Veff_tphi}. 
For the low-spin case ($\chi=0.3$), all perturbations decay, consistent with the absence of a trapping region. 
For the high-spin case ($\chi=0.7$), sufficiently large amplitudes lead to sustained growth and saturation of the scalar field, demonstrating the onset of nonlinear scalarization. 
This confirms that scalarization in the present model is triggered by a combination of rotation and nonlinear effects, rather than by a linear instability.

We have performed additional simulations with different parameter choices and numerical settings, and find that whenever condensation occurs, the late-time scalar field reaches a finite amplitude near the polar, near-horizon region that is of order $10^{-1}$, and often around $0.25$. 
This is of the same order as the largest values of $\max(\phi_H)$ in the corresponding solutions with full backreaction.
The decoupled evolutions therefore offer a useful physical interpretation of the scalarized black hole solutions constructed in the next section.

The results of this section suggest the following intuitive mechanism for spin-induced nonlinear scalarization. 
As the spin of a Kerr black hole increases, the Gauss-Bonnet invariant develops a negative near-horizon region in the polar direction. 
This region acts as an effective potential well for the scalar field, allowing sufficiently large perturbations to be trapped and amplified through the nonlinear coupling. 
Since the linearized scalar equation remains stable, this growth does not arise from a tachyonic instability but instead requires a finite-amplitude perturbation. 
Below a critical spin $\chi=0.5$, no such trapping region exists and the scalar field decays. 
Above this threshold, nonlinear effects can sustain a nontrivial scalar configuration, which then leads to scalarized black hole solutions in the system with full backreaction.

\section{Numerical setup}

\subsection{Metric ansatz}

To construct rotating scalarized black hole solutions with full backreaction, we solve the coupled Einstein-scalar-Gauss-Bonnet field equations using the stationary and axisymmetric Lewis-Papapetrou ansatz
\begin{eqnarray}
ds^2 = -F_0 dt^2 + F_1 \left( dr^2 + r^2 d\theta^2 \right)
+ r^2 \sin^2\theta \, F_2 \left( d\varphi - \frac{W}{r} dt \right)^2 ,
\end{eqnarray}
where the metric functions $F_0$, $F_1$, $F_2$ and $W$ depend on $r$ and $\theta$ only, and the scalar field is taken as $\phi=\phi(r,\theta)$. Following Ref.~\cite{Kunz:2019bhm} it is convenient to use the exponential parametrization
\begin{eqnarray}
F_0 =\left( \frac{1-r_h/r}{1+r_h/r} \right)^2 e^{f_0} , \qquad
F_1 = \left( 1+\frac{r_h}{r} \right)^4 e^{f_1} , \qquad
F_2 = \left( 1+\frac{r_h}{r} \right)^4 e^{f_2} .
\end{eqnarray}
This parametrization fixes the horizon at $r=r_h$ and factors out the leading near-horizon behavior of the metric, which simplifies the numerical implementation and allows one to impose regular boundary conditions on the functions $f_0$, $f_1$, and $f_2$.

\subsection{Boundary conditions}

The boundary conditions are chosen to impose asymptotic flatness, regularity at the event horizon and on the symmetry axis, and equatorial reflection symmetry of the even-parity solutions considered here. 
Asymptotic flatness and vanishing scalar field require
\begin{eqnarray}
f_0 = f_1 = f_2 = W = \phi = 0
\qquad \text{as } r\to\infty .
\end{eqnarray} 
Regularity at the horizon implies
\begin{eqnarray}
\partial_r f_0 = \partial_r f_1 = \partial_r f_2 = \partial_r \phi = 0
\qquad \text{at } r=r_h ,
\end{eqnarray}
together with a constant horizon angular velocity,
\begin{eqnarray}
W = r_h \Omega_H
\qquad \text{at } r=r_h .
\end{eqnarray}
These conditions guarantee that the horizon is regular and corresponds to a Killing horizon with constant angular velocity. 
Regularity on the symmetry axis requires
\begin{eqnarray}
\partial_\theta f_0 = \partial_\theta f_1 = \partial_\theta f_2 = \partial_\theta W = \partial_\theta \phi = 0
\qquad \text{at } \theta=0 ,
\end{eqnarray}
and the absence of conical singularities implies
\begin{eqnarray}
F_1 = F_2
\qquad \text{at } \theta=0 .
\end{eqnarray}
Equivalently, this condition gives $f_1=f_2$ on the axis.

For the even-parity scalarized branch considered here, reflection symmetry implies
\begin{eqnarray}
\partial_\theta f_0 = \partial_\theta f_1 = \partial_\theta f_2 = \partial_\theta W = \partial_\theta \phi = 0
\qquad \text{at } \theta=\frac{\pi}{2} .
\end{eqnarray}
The reflection symmetry imposed at the equatorial plane restricts the analysis to the even-parity sector, which is sufficient to capture the scalarized solutions considered in this work (see, e.g.,~\cite{Berti:2020kgk} for parity-odd spin-induced scalarized solutions).

\subsection{Extraction of physical quantities}

The ADM mass, angular momentum, and scalar charge are obtained from asymptotic fits of the metric components and the scalar field,
\begin{eqnarray}
g_{tt} = -1 + \frac{2M}{r} +  \cdots,
\qquad
g_{t\varphi} = -\frac{2J}{r} \sin^2\theta + \cdots,
\qquad
\phi(r) = \frac{Q_s}{r} + \cdots .
\end{eqnarray}
In practice, the fits are performed on the outermost finite radial points, excluding the compactified boundary $x=1$. 
We have verified that the extracted quantities are insensitive to the precise fitting region.
The dimensionless spin is defined by
\begin{eqnarray}
\chi \equiv \frac{J}{M^2} .
\end{eqnarray}
This definition allows for a direct comparison with Kerr black holes.

The horizon angular velocity and the Hawking temperature are obtained from the horizon data,
\begin{eqnarray}
\Omega_H = \frac{W(r_h,\theta)}{r_h} ,
\end{eqnarray}
\begin{eqnarray}
T_H = \frac{1}{16\pi r_h}
\exp\left[ \frac{f_0(r_h,\theta)-f_1(r_h,\theta)}{2} \right] .
\end{eqnarray}
For a physically consistent solution, both quantities must be independent of $\theta$. 
In our numerical solutions, any residual angular dependence is used as a diagnostic of numerical accuracy. 

The horizon area is extracted from the induced metric on the horizon as
\begin{eqnarray}
A_H = 32\pi r_h^2 \int_0^{\pi} d\theta \, \sin\theta 
\exp\left[ \frac{f_1(r_h,\theta)+f_2(r_h,\theta)}{2} \right].
\end{eqnarray}
The entropy is computed from the Wald formula,
\begin{eqnarray}
S_H = \frac{A_H}{4} + \frac{\alpha}{2} \int_H d^2x\,\sqrt{h}\, \zeta(\phi) \, R^{(2)} ,
\end{eqnarray}
where $h$ is the determinant of the induced metric on the horizon section and $R^{(2)}$ is its Ricci scalar.
 
In the following, we present all physical quantities in dimensionless form, which allows for a direct comparison between different solution families and highlights the scaling properties of the system.
The dimensionless quantities are denoted by hatted symbols. 
Using the Gauss-Bonnet coupling constant $\alpha$, we introduce the characteristic length scale $\sqrt{|\alpha|}$ and define
\begin{eqnarray}
\hat{r}_h = \frac{r_h}{\sqrt{|\alpha|}} , \qquad
\hat{M} = \frac{M}{\sqrt{|\alpha|}} , \qquad
\hat{Q}_s = \frac{Q_s}{\sqrt{|\alpha|}} ,
\end{eqnarray}
\begin{eqnarray}
\hat{\Omega}_H = \Omega_H \sqrt{|\alpha|} , \qquad
\hat{T}_H = T_H \sqrt{|\alpha|} ,
\end{eqnarray}
\begin{eqnarray}
\hat{A}_H = \frac{A_H}{|\alpha|} , \qquad
\hat{J} = \frac{J}{|\alpha|} , \qquad
\hat{S}_H = \frac{S_H}{|\alpha|} .
\end{eqnarray}

\subsection{Numerical method}

The field equations are solved as a coupled elliptic boundary-value problem on a two-dimensional grid in $(x,\theta)$. 
In the radial direction we use a finite-difference discretization on the compactified coordinate $x$:
\begin{eqnarray}
x = \frac{r-r_h}{r+L} ,
\end{eqnarray}
which maps the semi-infinite region $r\in[r_h,\infty)$ to the unit interval $x\in[0,1]$. 
The compactification parameter $L$ is a free parameter that controls the radial compactification.
It is chosen to provide an adequate resolution both near the horizon and in the asymptotic region.
We have checked that our results are insensitive to its precise value.
In the angular direction we use a pseudo-spectral representation adapted to the symmetry of the problem. 
For the even-parity sector this automatically enforces the regularity and reflection conditions at the symmetry axis and at the equatorial plane at the level of the functional basis.

After discretization, the nonlinear algebraic system is solved by a Newton-Raphson iteration. 
In most calculations reported below, we use a $601\times 16$ grid.
The residuals of the field equations are typically smaller than $\sim 10^{-10}$. 
We have performed convergence tests with different grid resolutions and verified that the physical quantities reported below are stable within the quoted numerical accuracy.

As a nontrivial consistency check of the numerical solutions, we verify the generalized Smarr relation for the scalarized black hole solutions
\begin{eqnarray}
M + M_s = 2 T_H S_H + 2 \Omega_H J,
\end{eqnarray}
where the contribution of the scalar field $M_s$ is given by
\begin{eqnarray}
M_s = \frac{\alpha}{8\pi}\int_{\Sigma} d^3x \sqrt{-g}\, \zeta(\phi)\,\mathcal{G} .
\end{eqnarray}
Accordingly, we define the relative Smarr error as
\begin{eqnarray}
\epsilon_{\rm Smarr} = \left| \frac{M + M_s - 2 T_H S_H - 2 \Omega_H J}{M + M_s} \right| .
\end{eqnarray}
For the scalarized black hole solutions discussed in this work, the relative error in the Smarr relation is typically of order $\sim 10^{-4}$ or smaller, confirming the overall consistency and accuracy of the numerical solutions.

\section{Scalarized black hole solutions}

We now present the properties of the scalarized black hole solutions with full backreaction and determine their domain of existence. 
As discussed above, in the present model scalarization does not originate from a linear instability, and the resulting solution space is therefore expected to differ qualitatively from the standard spontaneous scalarization scenario.

\begin{figure}[htbp]
    \centering
    \includegraphics[width=0.5\textwidth]{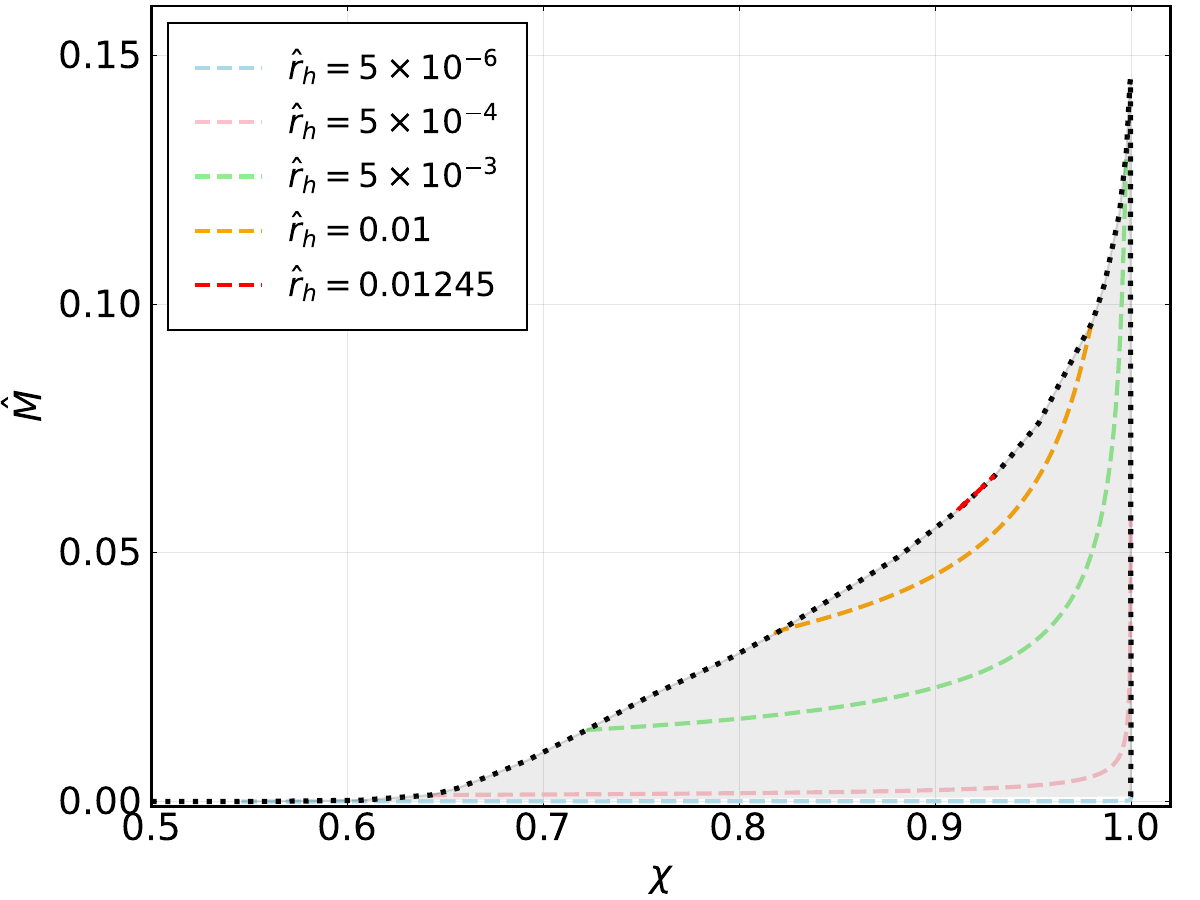}
    \caption{Domain of existence of scalarized black holes in the $(\chi,\hat M)$ plane. Colored curves show several representative fixed-$\hat r_h$ families, while the dotted boundary indicates the full numerically explored existence region. The solutions occupy a finite high-spin wedge. The low-spin boundary moves toward $\chi=0.5$, whereas toward the high-spin end the domain rises steeply and extends along a boundary close to $\chi=1$.}
    \label{fig_chi_M}
\end{figure}

We begin with the domain of existence of the scalarized solutions in the $(\chi,\hat M)$ plane, shown in Fig.~\ref{fig_chi_M}. 
The solutions exhibit a finite high-spin wedge, rather than a narrow band associated with a linear bifurcation line. 
In this sense, the present case differs from the spin-induced spontaneous scalarization of Ref.~\cite{Herdeiro:2020wei,Berti:2020kgk}, where the hairy branch is tied to the onset of a tachyonic instability on the Kerr background.
Instead, the present branch is closer to the nonlinear scalarization discussed in Refs.~\cite{Doneva:2021tvn,Lai:2023gwe}, where scalarized solutions need not arise from a linear zero mode.

Across the different branches, the minimum spin required for scalarization decreases systematically and tends toward $\chi=0.5$. 
In our present data, the smallest family reaches $\hat r_h=10^{-8}$, for which the minimum spin is already as low as $\chi\simeq0.503$. 
At the same time, the scalar charge drops to the level of $10^{-15}$ and the maximum horizon scalar field to the level of $10^{-6}$, indicating that the branch approaches $\chi=0.5$ while the hair tends to vanish. 
By contrast, toward the high-spin end the maximal allowed mass increases rapidly, so that the domain opens up significantly as $\chi$ approaches 1. 
These two trends produce the wedge-shaped high-spin region in the $(\chi,\hat M)$ plane.

The value $\chi=0.5$ should not be interpreted here as a linear onset in the sense of spontaneous scalarization. 
In the present model, the coupling satisfies $\zeta''(0)=0$, so that the linearized scalar equation on Kerr reduces to the massless wave equation and no linear tachyonic onset is present. 
The approach to $\chi=0.5$ is therefore more naturally understood as a geometric threshold associated with the appearance of a trapping region near the poles, consistent with the decoupled dynamics discussed in Sec. III.

\begin{figure}[htbp]
\centering
\subfloat[$\hat Q_s$ as a function of $\hat M$\label{fig_M_Qs_two_branches}]{
    \includegraphics[width=0.45\textwidth]{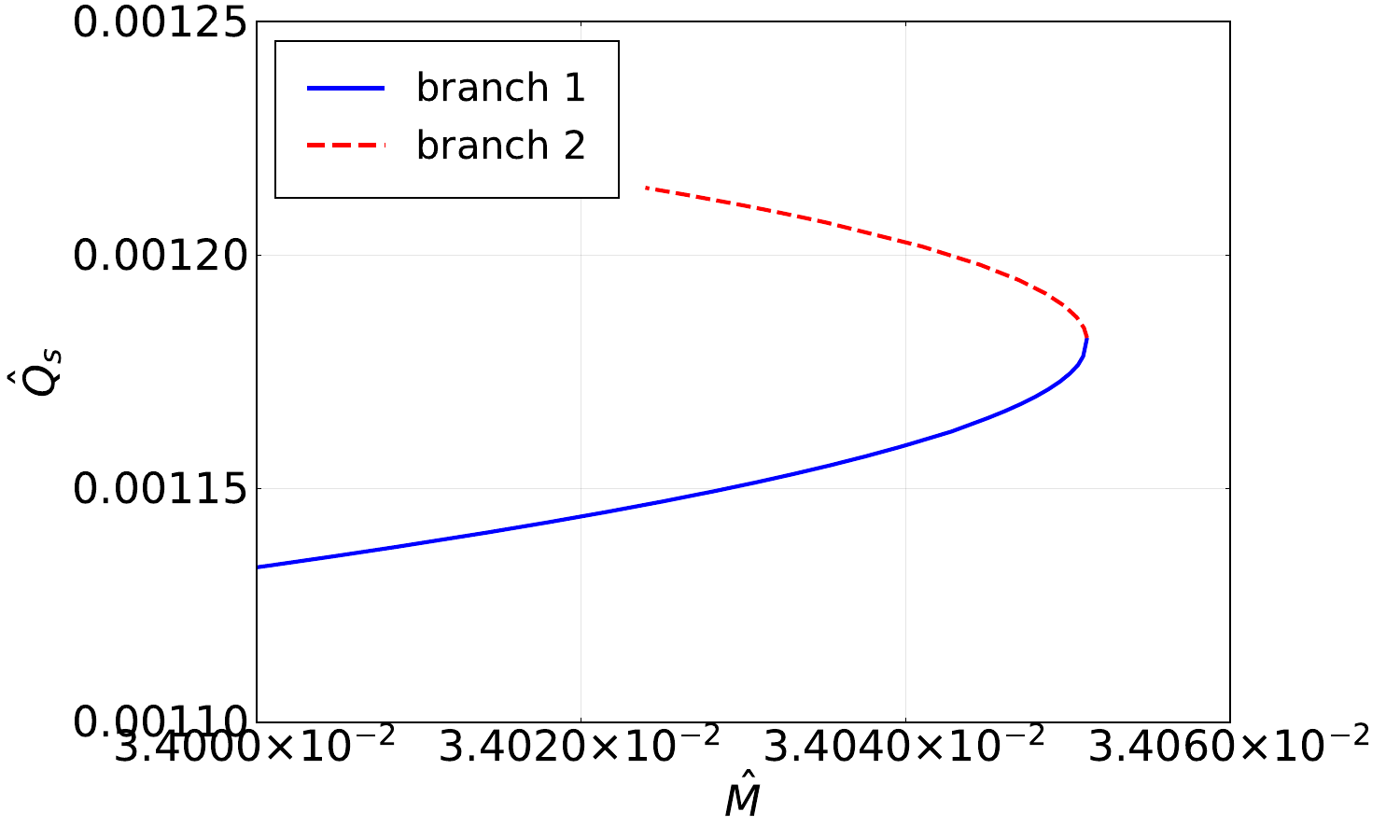}
}
\hfill
\subfloat[Entropy difference between the two branches\label{fig_M_dSh_two_branches}]{
    \includegraphics[width=0.45\textwidth]{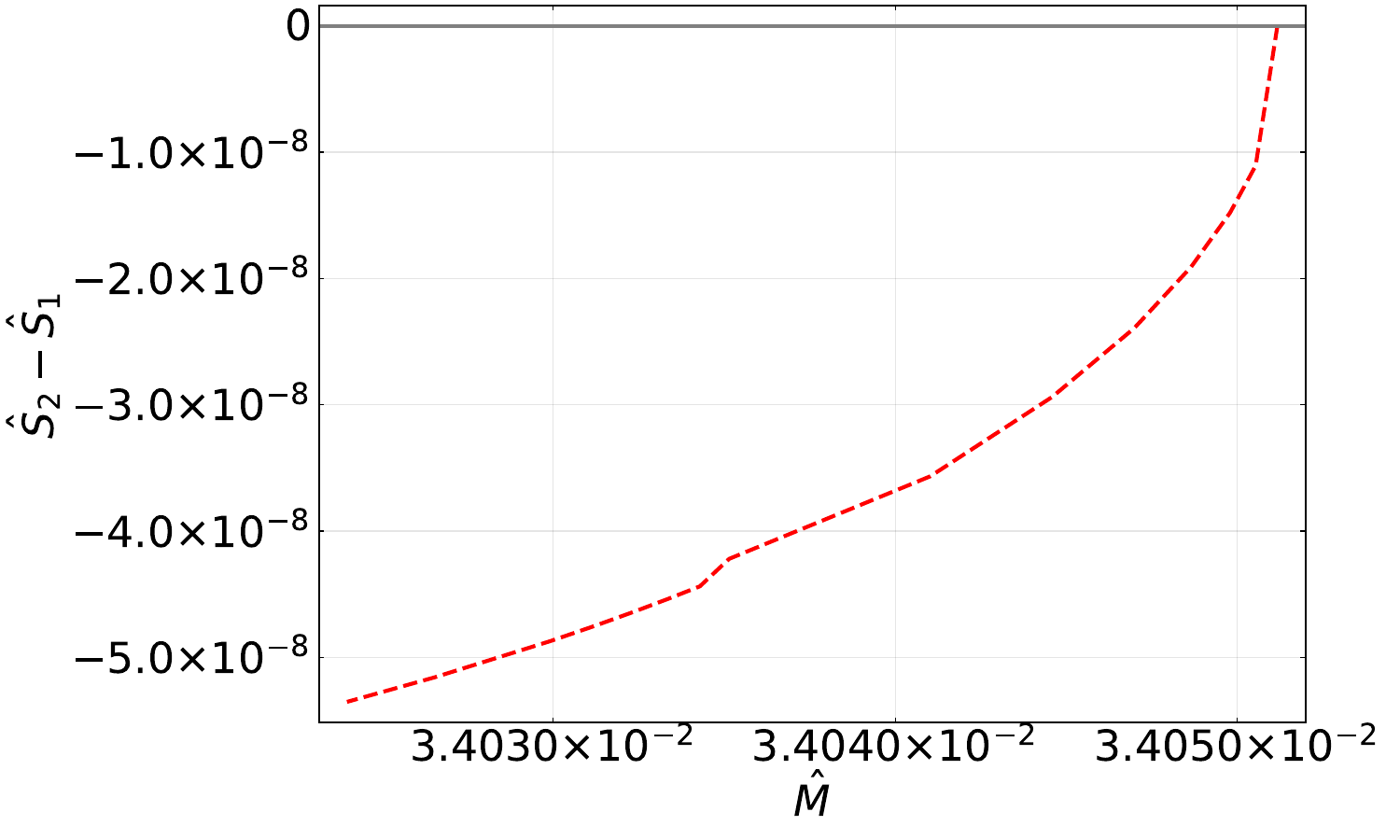}
}
\caption{
Two-branch structure of the scalarized solutions for $\chi=0.8185$, obtained by using $\hat r_h$ as the continuation parameter. 
Panel (a) shows $\hat Q_s$ as a function of $\hat M$, where the two portions are denoted as branch 1 and branch 2. 
Panel (b) shows the entropy difference $\hat S_2-\hat S_1$ between the two branches at the same $\hat M$ and $\chi$.
}
\label{fig_fixed_chi}
\end{figure}

To further clarify the internal structure of the scalarized domain, it is useful to examine the solutions at fixed dimensionless spin. 
As shown in Fig.~\ref{fig_fixed_chi}(a), the curve\footnote{
The fixed-$\chi$ curve was constructed using the publicly available code associated with Ref.~\cite{Fernandes:2022gde}.
} exhibits a two-branch-like structure in the $(\hat M,\hat Q_s)$ plane, with a very short second branch. 
In our continuation, $\hat r_h$ increases continuously from branch 1 to branch 2. 
This suggests a continuous turning structure, similar to the turning behavior found in certain parameter regions of scalarized black hole solutions in the literature~\cite{Silva:2018qhn,Herdeiro:2026sur}. 
This behavior differs from the static nonlinear scalarization case, where $\hat r_h$ typically reaches a maximum at the branching point. 
This type of turning behavior is often associated with changes in stability and may indicate a nontrivial internal structure of the scalarized branch, although a detailed stability analysis is beyond the scope of the present work.

Fig.~\ref{fig_fixed_chi}(b) shows the entropy difference between the two scalarized branches at the same $\hat M$ and $\chi$, where $\hat S_1$ and $\hat S_2$ correspond to branch 1 and branch 2, respectively.
The plotted quantity $\hat S_2-\hat S_1$ remains negative over the overlapping mass range. 
Thus, among the two scalarized solutions with the same $\hat M$ and $\chi$, the short second branch has lower entropy and is thermodynamically disfavored with respect to the main branch.

\begin{figure}[htbp]
    \centering
    \includegraphics[width=0.5\textwidth]{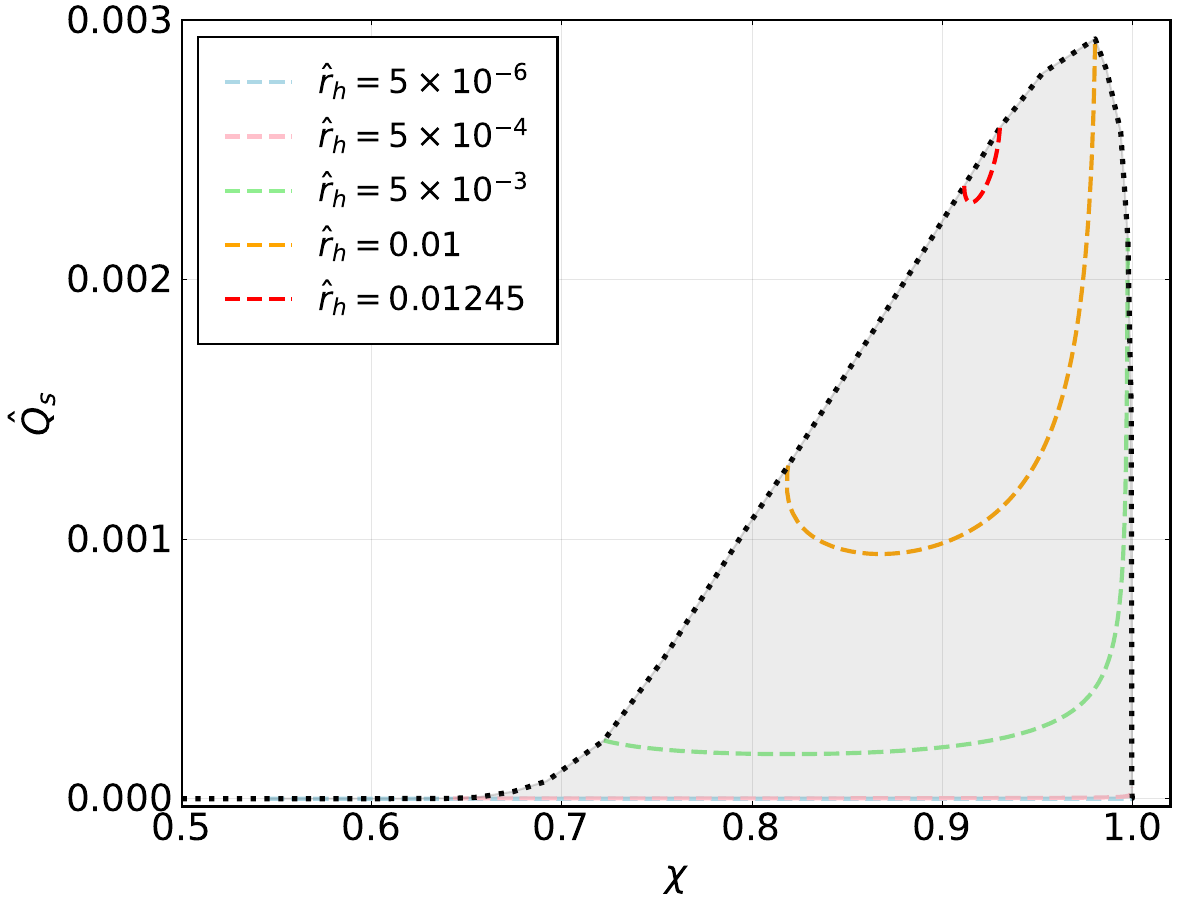}
    \caption{Scalar charge in the $(\chi,\hat Q_s)$ plane. Colored curves show several representative fixed-$\hat r_h$ families, while the dotted boundary indicates the full numerically explored existence region. Larger scalar charge is concentrated near the high-spin end, whereas the low-spin side is associated with much weaker hair.}
    \label{fig_chi_Qs}
\end{figure}

We next turn to the global behavior of the scalar charge over the scalarized domain, shown in Fig.~\ref{fig_chi_Qs}. 
In general, the scalar charge becomes larger toward the high-spin end, indicating that stronger rotation is accompanied by stronger scalar hair. 
This correlation further supports the interpretation that rotation acts as the primary driver of scalarization in the present model.
At the same time, the branches that extend to lower values of $\chi$ carry much smaller scalar charge. 
In particular, although smaller $\hat r_h$ allows the solutions to approach $\chi=0.5$ more closely, the overall magnitude of $\hat Q_s$ is then significantly reduced. 
Thus, the approach to the low-spin boundary is also an approach to a weak-hair regime.

\begin{figure}[htbp]
    \centering
    \includegraphics[width=0.5\textwidth]{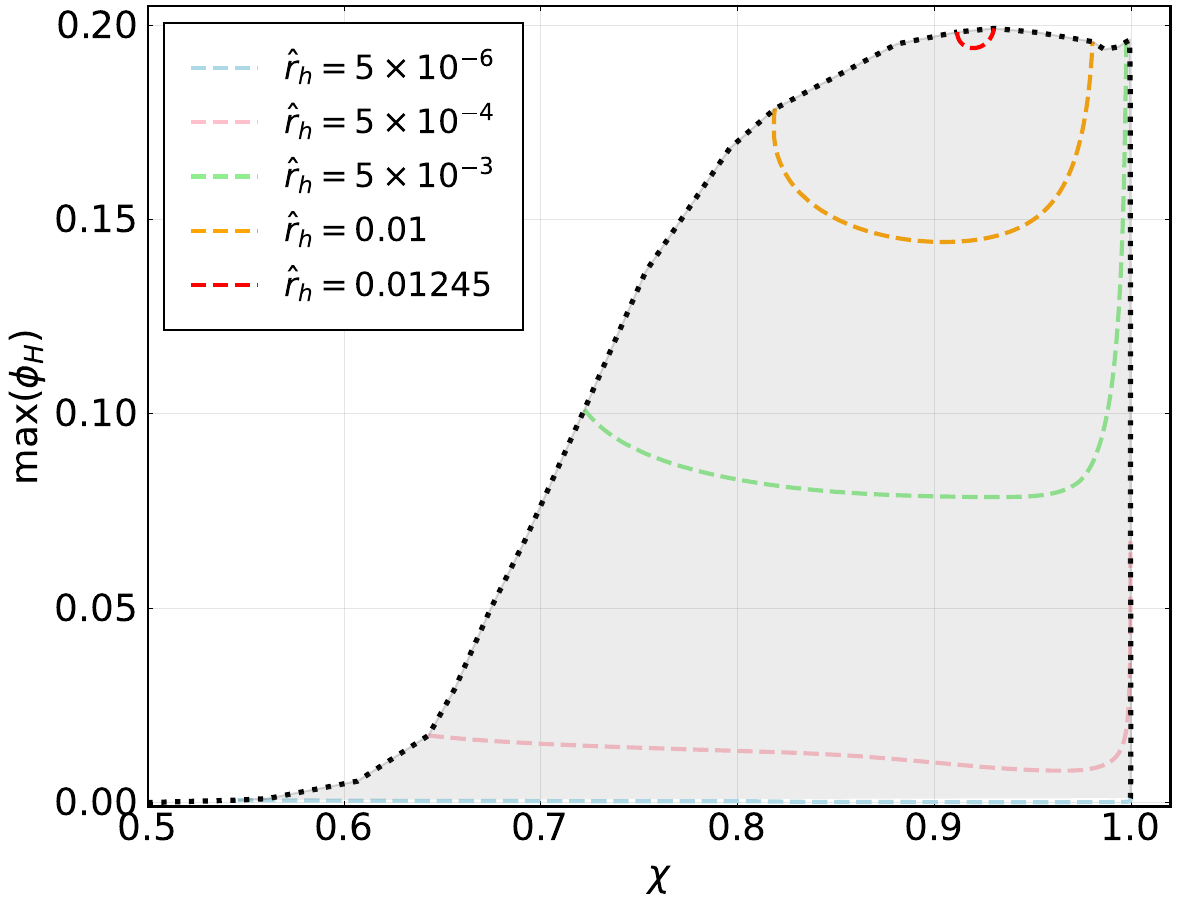}
    \caption{Maximum horizon scalar field in the $(\chi,\max(\phi_H))$ plane. Colored curves show several representative fixed-$\hat r_h$ families, while the dotted boundary indicates the full numerically explored existence region. Larger values of $\max(\phi_H)$ are concentrated near the high-spin end, whereas the low-spin side is associated with much weaker horizon scalarization.}
    \label{fig_chi_PhiH}
\end{figure}

A similar pattern is seen in the $(\chi,\max(\phi_H))$ plane, shown in Fig.~\ref{fig_chi_PhiH}. 
The horizon scalar field becomes much larger toward the high-spin end, showing that the near-horizon scalarization is strongest there, consistent with the expectation that the polar trapping region becomes most effective at high spin.
By contrast, as the branches extend toward lower values of $\chi$, $\max(\phi_H)$ decreases rapidly and approaches very small values. 
In particular, although smaller $\hat r_h$ allows the solutions to reach closer to $\chi=0.5$, the corresponding horizon scalar field is then strongly suppressed, consistent with the approach to a vanishing-hair limit. 
Thus, the low-spin side of the domain is again associated with a weak-hair regime. 
Representative maximum values are of order $10^{-1}$, reaching about $0.2$, which is comparable in magnitude to the saturation amplitude found in the decoupled time evolutions.

\begin{figure}[htbp]
    \centering
    \includegraphics[width=0.5\textwidth]{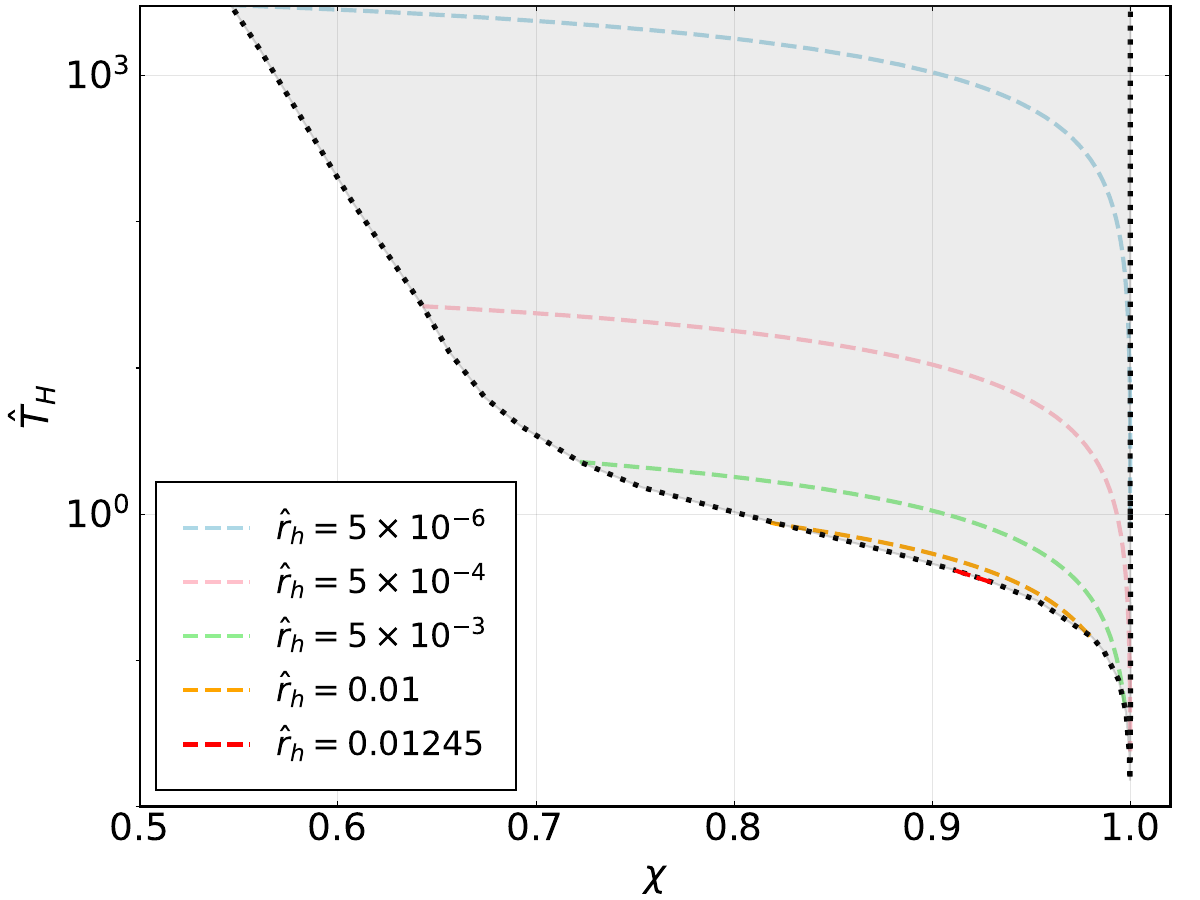}
    \caption{Temperature in the $(\chi,\hat T_H)$ plane. Colored curves show several representative fixed-$\hat r_h$ families, while the dotted boundary indicates the full numerically explored existence region. The scalarized branches move toward lower temperatures as $\chi$ approaches $1$, whereas the lower-spin side is associated with higher temperatures.}
    \label{fig_chi_TH}
\end{figure}

The behavior of the temperature is shown in the $(\chi,\hat T_H)$ plane in Fig.~\ref{fig_chi_TH}. 
In general, $\hat T_H$ decreases toward the high-spin end, so that the scalarized branches bend toward a low-temperature region as $\chi$ approaches $1$. 
This is consistent with the interpretation that the right edge of the domain is associated with a near-extremal regime. 
This behavior suggests that nonlinear scalarization is particularly efficient in rapidly rotating, low-temperature configurations.
At the same time, the branches that extend to smaller values of $\chi$ lie at higher temperatures, and this trend is more pronounced for smaller $\hat r_h$. 
Thus, the temperature plot provides a complementary view of the same domain structure: the low-spin side is connected to a high-temperature weak-hair regime, while the high-spin side approaches a low-temperature near-extremal edge.

\begin{figure}[htbp]
\centering
\subfloat[$\phi(\hat z,\hat\rho)$\label{fig:phi1}]{
    \includegraphics[width=0.45\textwidth]{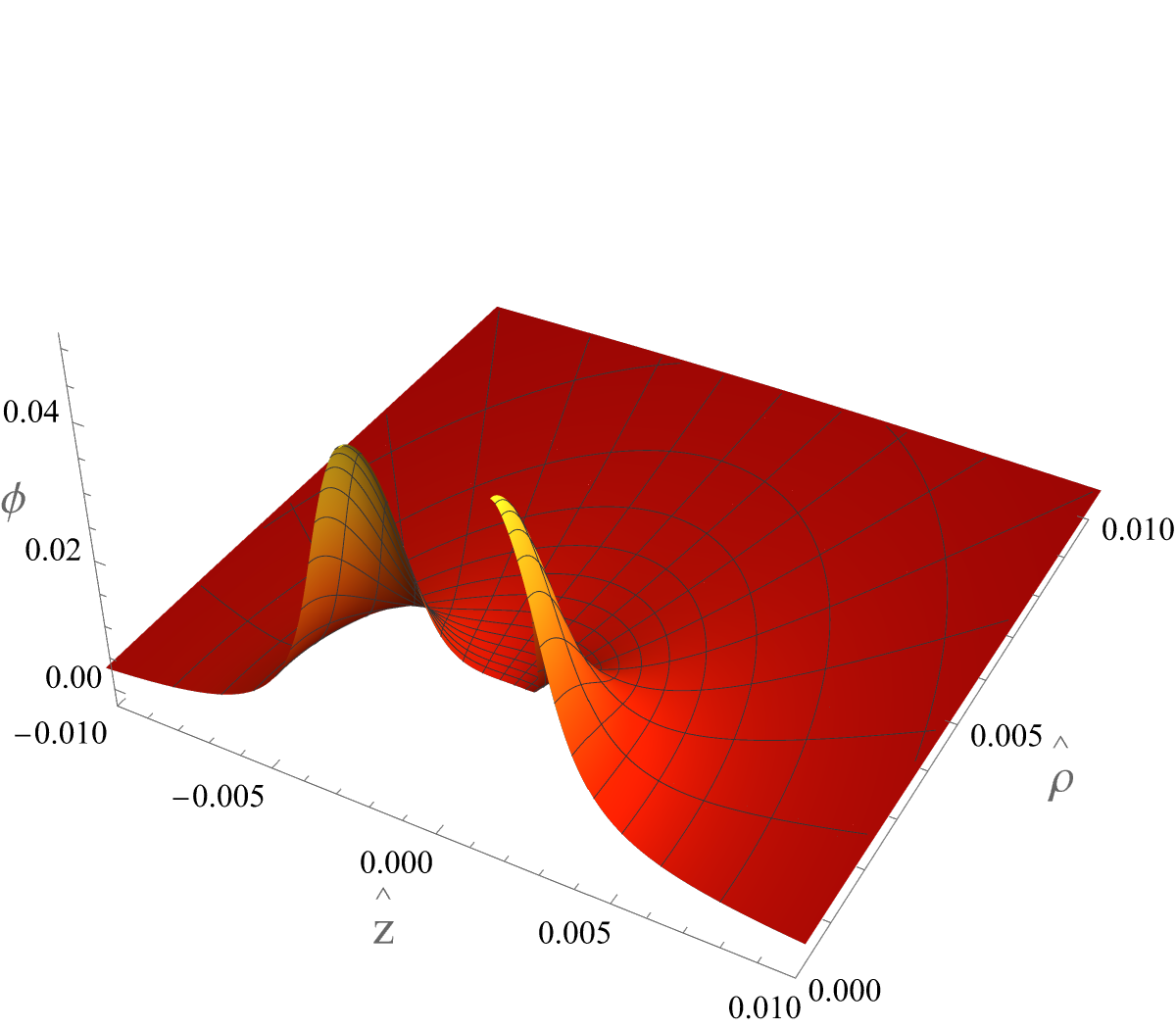}
}
\hfill
\subfloat[Dimensionless Ricci scalar $\hat R(\hat z,\hat\rho)$\label{fig:ricci1}]{
    \includegraphics[width=0.45\textwidth]{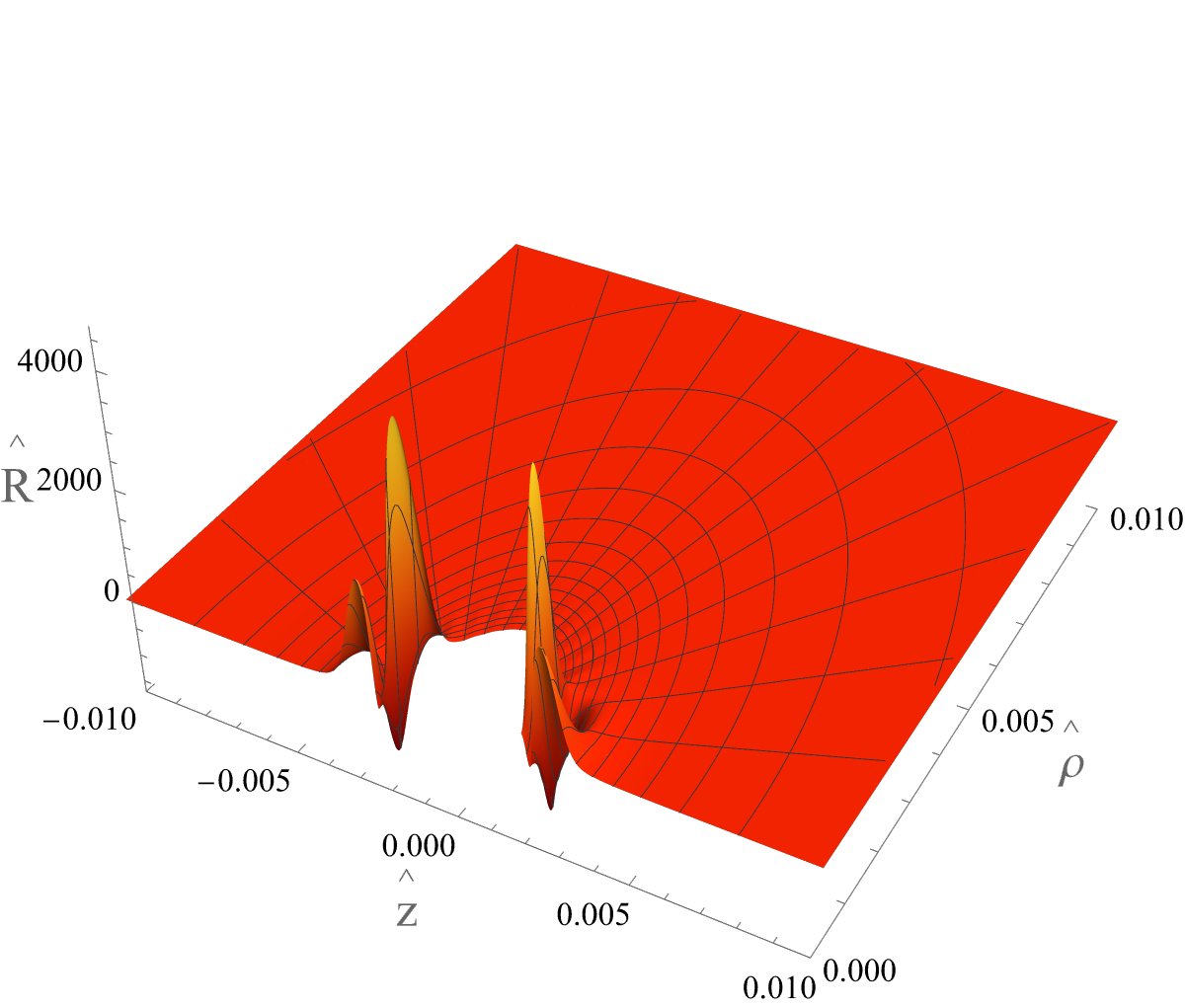}
}
\caption{Meridional profiles for a scalarized solution with $\hat r_h=0.002$ near the low-spin end of the branch, at $\chi\simeq0.677$. The scalar field is weak and mainly concentrated near the poles, while the Ricci scalar shows sharp localized peaks near the horizon.}
\label{fig_low_spin}
\end{figure}

\begin{figure}[htbp]
\centering
\subfloat[$\phi(\hat z,\hat\rho)$\label{fig:phi2}]{
    \includegraphics[width=0.45\textwidth]{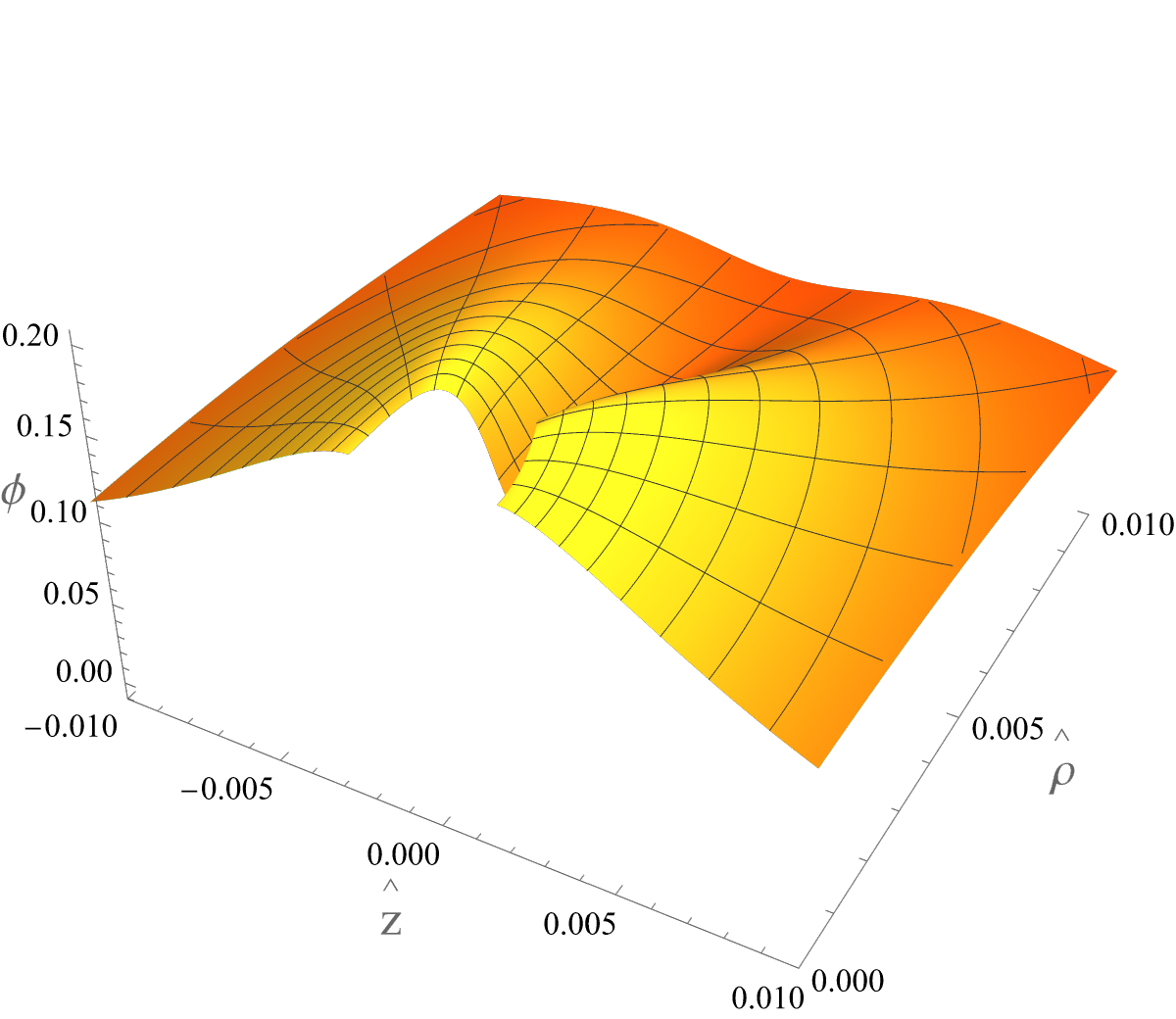}
}
\hfill
\subfloat[Dimensionless Ricci scalar $\hat R(\hat z,\hat\rho)$\label{fig:ricci2}]{
    \includegraphics[width=0.45\textwidth]{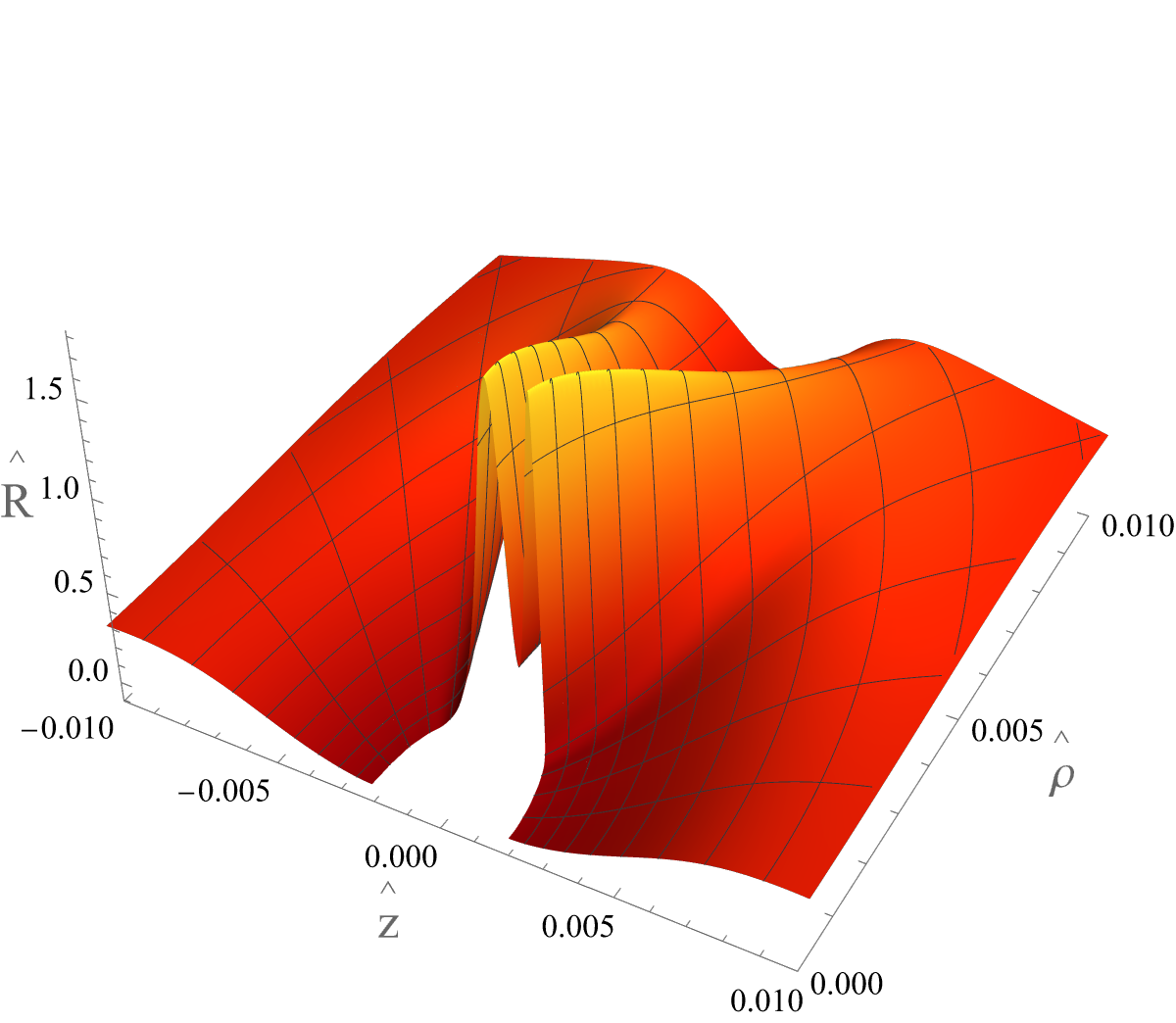}
}
\caption{Meridional profiles for a scalarized solution with $\hat r_h=0.002$ near the high-spin end of the branch, at $\chi\simeq0.999$. The scalar field becomes much stronger and remains concentrated near the poles. The Ricci scalar varies strongly near the horizon and its overall magnitude is much smaller than in the low-spin case.}
\label{fig_high_spin}
\end{figure}

The difference between the low-spin and high-spin ends of the branch can be seen more directly from the spatial profiles of the scalar field and the Ricci scalar. 
To visualize the spatial structure of the scalarized solutions, we consider meridional profiles in the dimensionless coordinates $\hat{\rho}=\hat r\sin\theta$ and $\hat z=\hat r\cos\theta$, where $\hat r=r/\sqrt{|\alpha|}$, together with the dimensionless Ricci scalar $\hat R=|\alpha|R$. 
In Figs.~\ref{fig_low_spin} and \ref{fig_high_spin}, we show $\phi(\hat z,\hat\rho)$ and the dimensionless Ricci scalar $\hat R(\hat z,\hat\rho)$ for the same family with $\hat r_h=0.002$, taking one solution near the low-spin end and another near the high-spin end. 
In both cases, the scalar field is mainly concentrated in the polar regions, in agreement with the effective-potential picture discussed above. 
Near the low-spin end, the scalar field is weak and the Ricci scalar develops sharp, localized peaks close to the horizon. 
Near the high-spin end, the scalar field becomes much stronger, while the Ricci scalar still shows pronounced variation near the horizon, but with a broader profile and a much smaller overall magnitude. 
These profiles illustrate how the branch evolves from a weak-hair configuration with strong localized curvature response to a strongly scalarized state with weak curvature.

Overall, these results provide a consistent picture in which spin-induced nonlinear scalarization is controlled by the interplay between the geometry of the Kerr background and the nonlinear scalar coupling. 
The emergence of a finite high-spin domain, together with the absence of a bifurcation line, highlights a qualitative departure from the standard spontaneous scalarization scenario. 
In particular, the solutions found here demonstrate that scalarized black holes can exist even when the Kerr background is linearly stable, provided that nonlinear effects are sufficiently strong.

\section{Summary and discussion}

In this work, we have investigated spin-induced scalarization of Kerr black holes in an Einstein-scalar-Gauss-Bonnet model that does not admit a linear tachyonic instability of the scalar-free solution.
This allowed us to isolate a genuinely nonlinear scalarization mechanism driven by rotation.

By analyzing the decoupled scalar dynamics on fixed Kerr backgrounds, we have shown that sufficiently rapid rotation modifies the Gauss-Bonnet invariant such that a negative near-horizon region develops in the polar direction. 
This region acts as a geometric trapping zone for the scalar field and provides the physical origin of the nonlinear scalar growth observed in our simulations. 
In particular, this mechanism becomes effective only above a threshold spin $\chi=0.5$, which should be understood as a geometric condition for scalar confinement rather than a linear instability threshold.

We have then constructed stationary scalarized black hole solutions with full backreaction and determined their domain of existence. 
In contrast to the spin-induced spontaneous scalarization scenario, where the scalarized solutions arise from a bifurcation line and form a narrow band, we find that the solutions occupy a finite high-spin wedge in the $(\chi, \hat{M})$ plane. 
Toward the high-spin end, the scalar hair becomes stronger and the solutions approach a near-extremal regime, while toward the low-spin boundary the scalar field is strongly suppressed, corresponding to a weak-hair limit as $\chi\to 0.5$.

These results demonstrate that scalarized black holes can exist even when the Kerr background is linearly stable, provided that nonlinear effects are sufficiently strong. 
In this sense, rotation acts as a purely geometric trigger for scalarization, and the existence of scalarized solutions cannot, in general, be inferred from linear perturbation theory alone.

An important open question concerns the dynamical stability of the scalarized solutions constructed here. 
Since the present branch is not connected to the Kerr solution through a linear instability, its stability properties may differ significantly from those of spontaneously scalarized black holes. 
A dedicated stability analysis, as well as fully nonlinear time evolutions including metric backreaction, would therefore be essential to assess the physical relevance of these solutions.

Finally, the existence of a finite high-spin domain suggests that nonlinear scalarization may play a role in the physics of rapidly rotating black holes in modified gravity theories. 
Exploring possible observational signatures, for example in gravitational-wave emission or black hole shadow properties, would be an interesting direction for future work.

% \hspace*{3em}
% \vspace{1cm}

\section*{Acknowledgments}
This work is supported by the National Natural Science Foundation of China (NSFC) under Grant Nos.~12305064, 12365009, 12565010, and 12205123. Y.S.M. is supported by the National Research Foundation of Korea (NRF) grant funded by the Korea government(MSIT) (RS-2022-NR069013).

% \vspace{1cm}

\appendix
\section{Entropy comparison with Kerr}\label{AppendixA}

As an auxiliary thermodynamic diagnostic, we evaluate the difference between the dimensionless Wald entropy of the scalarized solutions and that of a Kerr black hole with the same $\hat M$ and $\chi$. 
This comparison provides a useful indication of whether the scalarized branch may be thermodynamically favored relative to Kerr.
More precisely, for each numerical solution we define
\begin{eqnarray}
\Delta \hat S = \hat S_{H}-\hat S_{\rm Kerr},
\end{eqnarray}
where $S_H$ is the Wald entropy of the scalarized black hole. 
A positive value of $\Delta \hat S$ would indicate that the scalarized solution has a higher entropy than the corresponding Kerr black hole and may therefore be thermodynamically preferred.
The Kerr reference entropy is computed at the same $\hat M$ and $\chi$ according to
\begin{eqnarray}
\hat S_{\rm Kerr}
= \frac{S_{\rm Kerr}}{|\alpha|}
= 2\pi \hat M^2 \left(1+\sqrt{1-\chi^2}\right).
\end{eqnarray}

\begin{figure}[htbp]
\centering
\subfloat[Global view\label{figchiS1}]{
    \includegraphics[width=0.45\textwidth]{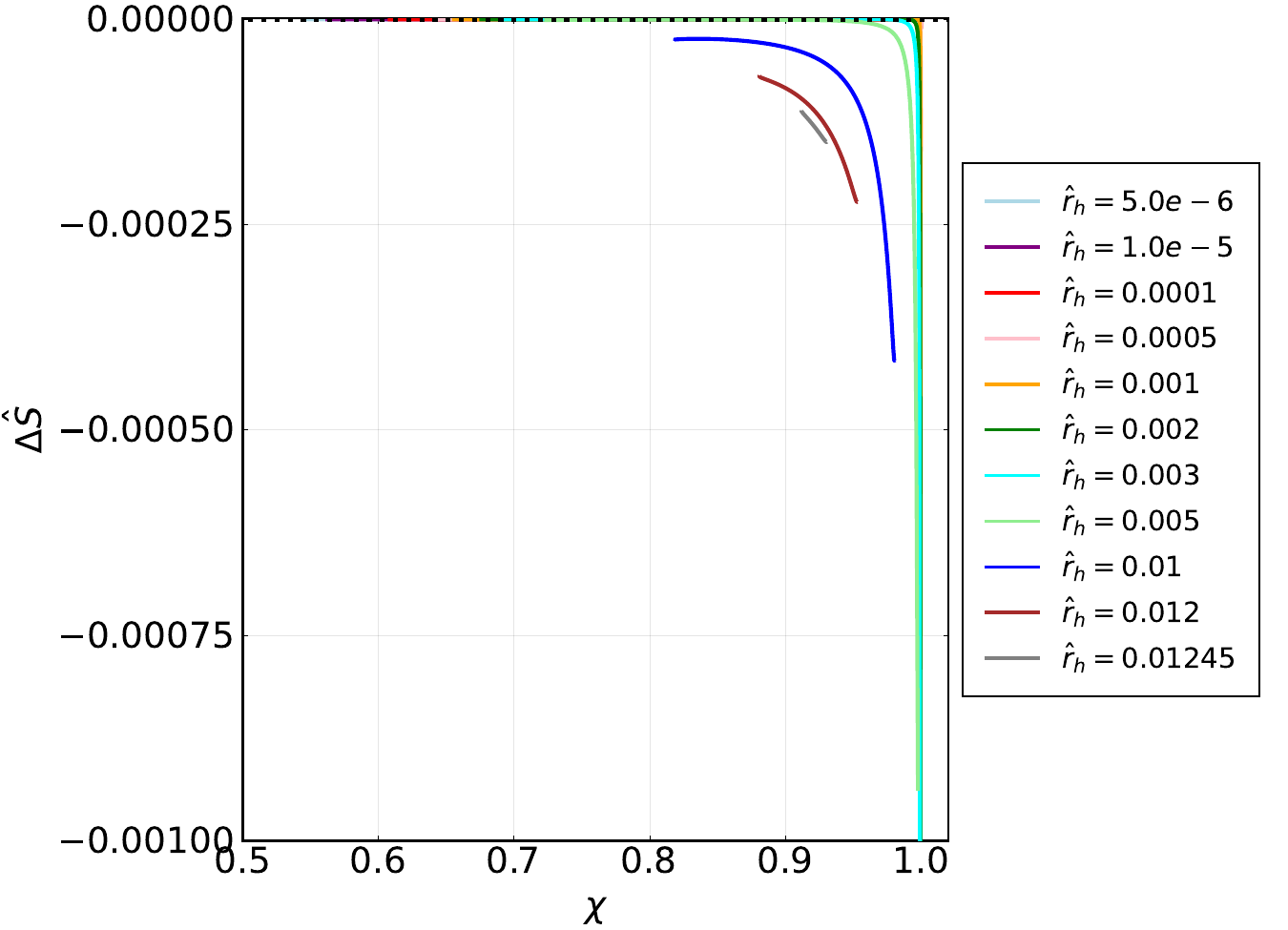}
}
\hfill
\subfloat[Details\label{figchiS2}]{
    \includegraphics[width=0.45\textwidth]{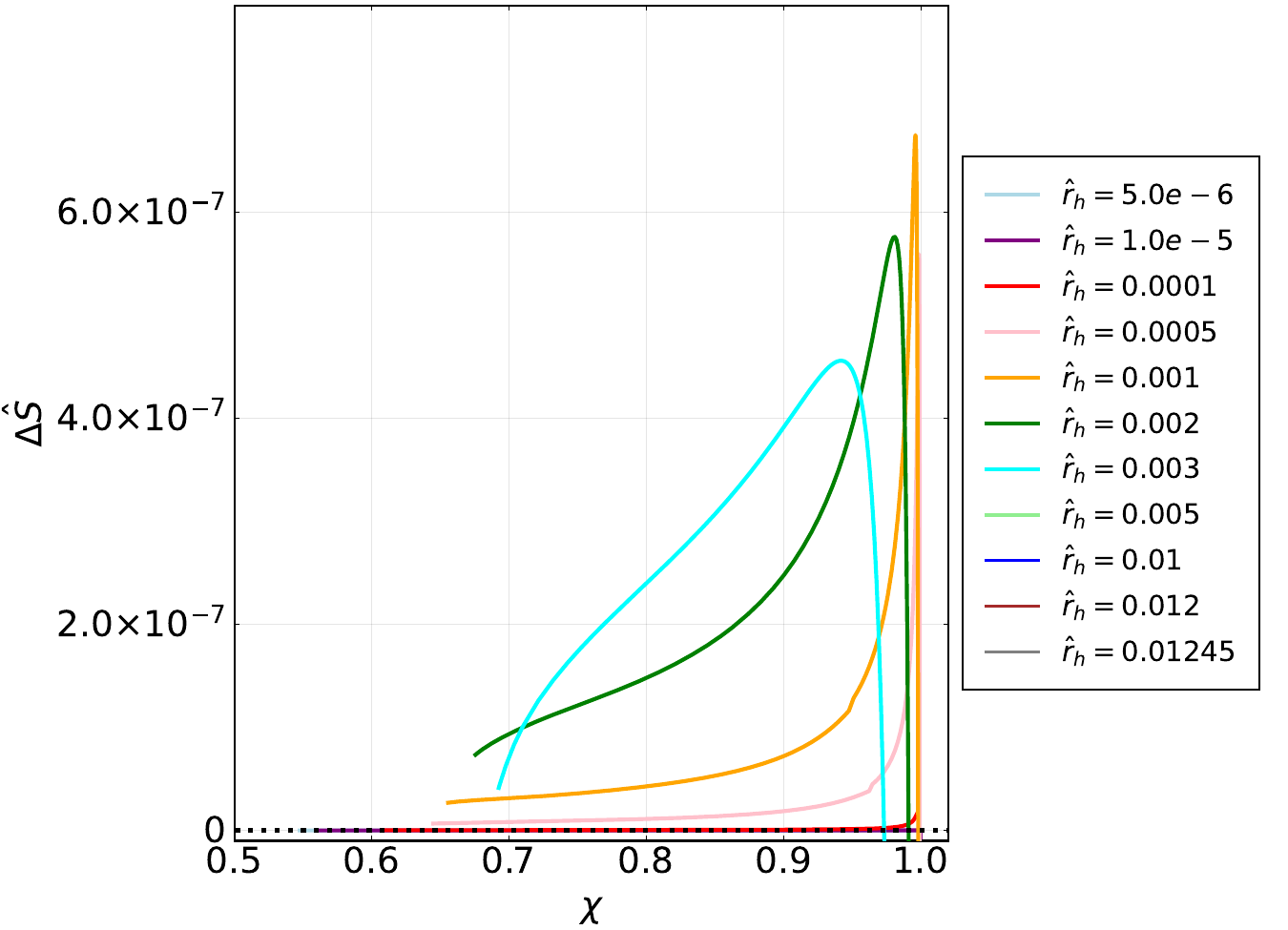}
}
\caption{
Entropy difference $\Delta\hat S=\hat S_H-\hat S_{\rm Kerr}$ as a function of $\chi$ for the fixed-$\hat r_h$ families, where the Kerr reference is taken at the same $\hat M$ and $\chi$. 
Panel (a) shows the global view, while panel (b) zooms in on the details where a small positive $\Delta\hat S$ can appear for some low-$\hat r_h$ branches.
}
\label{fig_chi_S}
\end{figure}

Figure \ref{fig_chi_S} shows $\Delta \hat S$ as a function of $\chi$ for all fixed-$\hat r_h$ families. 
The entropy difference is numerically very small across the domain of existence. 
The global view illustrates the overall scale of $\Delta \hat S$, while the zoomed-in panel resolves a narrow region near the high-spin end where a small positive $\Delta \hat S$ may appear for some low-$\hat r_h$ branches. 
This suggests that, in a limited parameter range, the scalarized solutions could exhibit a slight thermodynamic preference over Kerr.
However, given the small magnitude of the effect, we regard this comparison as indicative rather than conclusive.

\section{Two-branch structure at fixed horizon radius}
\label{AppendixB}

In this appendix, we provide additional evidence for the nontrivial internal structure of the scalarized domain. 
The main text discussed a fixed-$\chi$ slice, where the scalarized solutions exhibit a short second branch in the $(\hat M,\hat Q_s)$ plane. 
Here we show that a similar branch-like behavior can also be observed when the solutions are followed along a fixed-$\hat r_h$ family.

\begin{figure}[htbp]
\centering
\subfloat[$\hat Q_s$ as a function of $\chi$\label{fig_chi_Qs_two_branches}]{
    \includegraphics[width=0.45\textwidth]{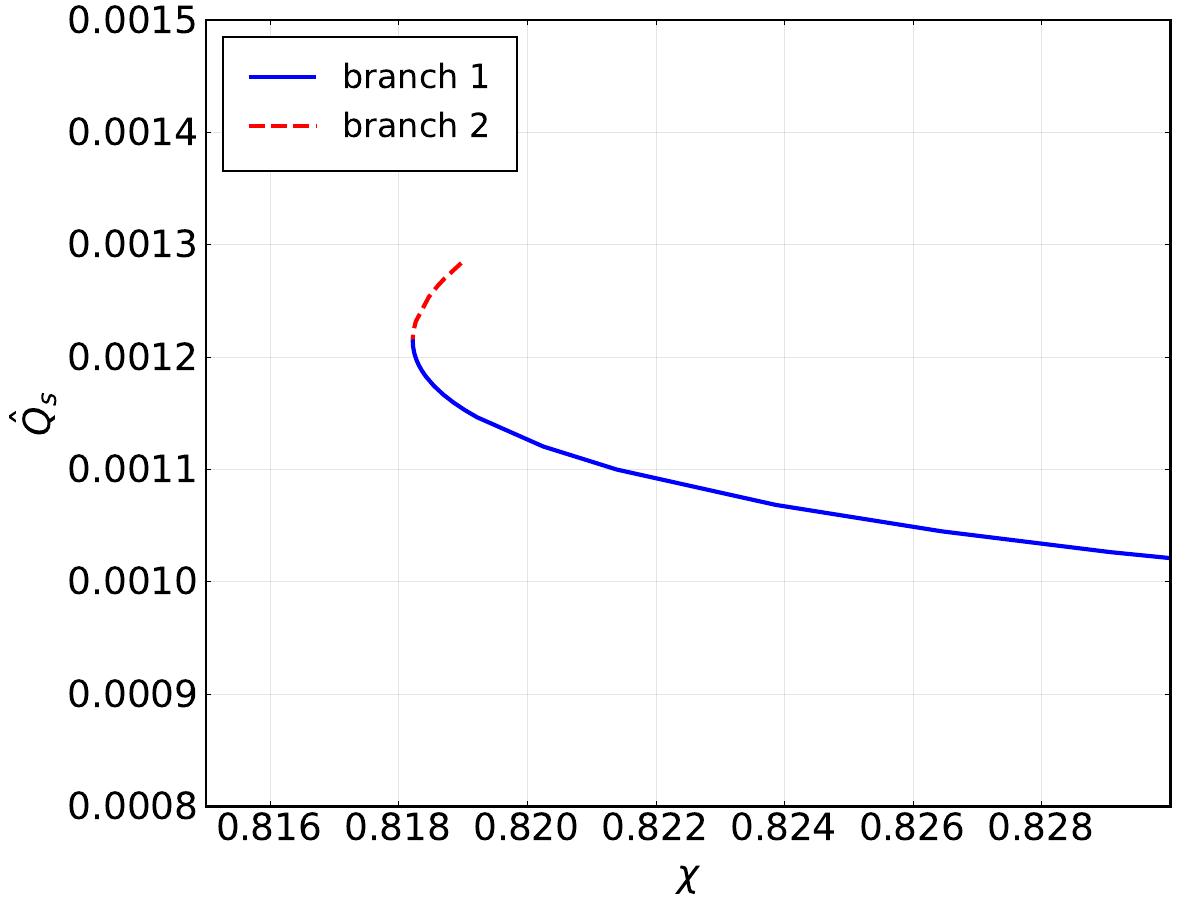}
}
\hfill
\subfloat[$\hat M$ as a function of $\chi$\label{fig_chi_M_two_branches}]{
    \includegraphics[width=0.45\textwidth]{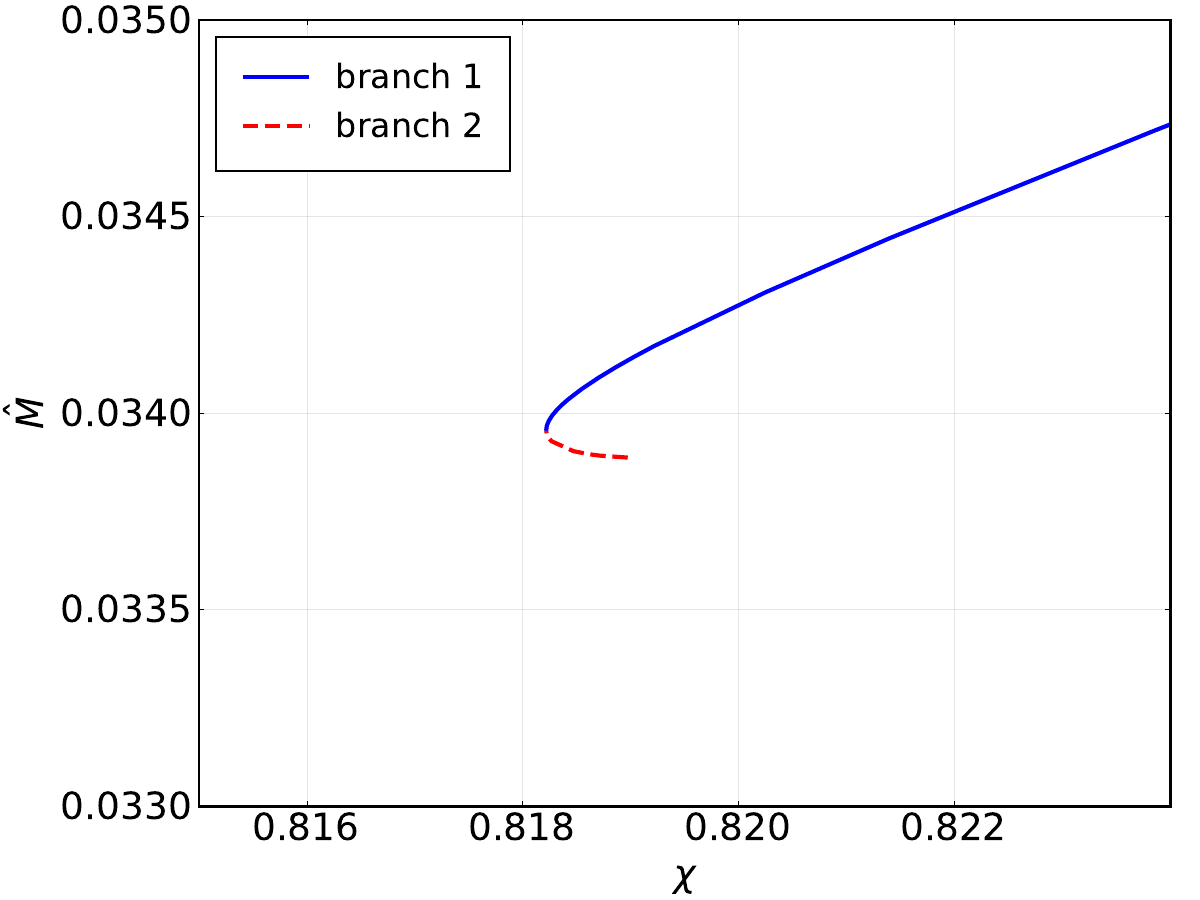}
}
\caption{
Zoomed-in view of the fixed-$\hat r_h=0.01$ family near the low-spin endpoint. 
Panel (a) shows the scalar charge $\hat Q_s$ as a function of the spin $\chi$, while panel (b) shows the mass $\hat M$ as a function of $\chi$. 
The two nearby portions of the curves, labeled as branch 1 and branch 2, indicate a local branch-like or turning structure in this narrow parameter region.
}
\label{fig_fixed_rh_two_branches}
\end{figure}

Figure~\ref{fig_fixed_rh_two_branches} shows a zoomed-in view of the fixed-$\hat r_h=0.01$ family near its low-spin endpoint. 
In the $(\chi,\hat Q_s)$ plane, the curve develops two nearby portions, labeled as branch 1 and branch 2. 
These two portions correspond to solutions with the same $\hat r_h$ and $\chi$, but with different scalar charges. 
The same structure is also seen in the $(\chi,\hat M)$ plane, where the two portions have slightly different masses. This behavior indicates that, for certain values of $\hat r_h$, the fixed-$\hat r_h$ families may develop a local fold or turning structure near the low-spin endpoint. 
This behavior further supports the interpretation that the scalarized domain possesses a nontrivial internal structure, potentially associated with turning points in the solution space. Such features are often linked to changes in stability and merit further investigation.

\clearpage

\bibliographystyle{apsrev4-1}
\bibliography{GR}

\begin{thebibliography}{41}%
\makeatletter
\providecommand \@ifxundefined [1]{%
 \@ifx{#1\undefined}
}%
\providecommand \@ifnum [1]{%
 \ifnum #1\expandafter \@firstoftwo
 \else \expandafter \@secondoftwo
 \fi
}%
\providecommand \@ifx [1]{%
 \ifx #1\expandafter \@firstoftwo
 \else \expandafter \@secondoftwo
 \fi
}%
\providecommand \natexlab [1]{#1}%
\providecommand \enquote  [1]{``#1''}%
\providecommand \bibnamefont  [1]{#1}%
\providecommand \bibfnamefont [1]{#1}%
\providecommand \citenamefont [1]{#1}%
\providecommand \href@noop [0]{\@secondoftwo}%
\providecommand \href [0]{\begingroup \@sanitize@url \@href}%
\providecommand \@href[1]{\@@startlink{#1}\@@href}%
\providecommand \@@href[1]{\endgroup#1\@@endlink}%
\providecommand \@sanitize@url [0]{\catcode `\\12\catcode `\$12\catcode `\&12\catcode `\#12\catcode `\^12\catcode `\_12\catcode `\%12\relax}%
\providecommand \@@startlink[1]{}%
\providecommand \@@endlink[0]{}%
\providecommand \url  [0]{\begingroup\@sanitize@url \@url }%
\providecommand \@url [1]{\endgroup\@href {#1}{\urlprefix }}%
\providecommand \urlprefix  [0]{URL }%
\providecommand \Eprint [0]{\href }%
\providecommand \doibase [0]{http://dx.doi.org/}%
\providecommand \selectlanguage [0]{\@gobble}%
\providecommand \bibinfo  [0]{\@secondoftwo}%
\providecommand \bibfield  [0]{\@secondoftwo}%
\providecommand \translation [1]{[#1]}%
\providecommand \BibitemOpen [0]{}%
\providecommand \bibitemStop [0]{}%
\providecommand \bibitemNoStop [0]{.\EOS\space}%
\providecommand \EOS [0]{\spacefactor3000\relax}%
\providecommand \BibitemShut  [1]{\csname bibitem#1\endcsname}%
\let\auto@bib@innerbib\@empty
%</preamble>
\bibitem [{\citenamefont {Abbott}\ \emph {et~al.}(2016)\citenamefont {Abbott} \emph {et~al.}}]{LIGOScientific:2016aoc}%
  \BibitemOpen
  \bibfield  {author} {\bibinfo {author} {\bibfnamefont {B.~P.}\ \bibnamefont {Abbott}} \emph {et~al.} (\bibinfo {collaboration} {LIGO Scientific, Virgo}),\ }\href {\doibase 10.1103/PhysRevLett.116.061102} {\bibfield  {journal} {\bibinfo  {journal} {Phys. Rev. Lett.}\ }\textbf {\bibinfo {volume} {116}},\ \bibinfo {pages} {061102} (\bibinfo {year} {2016})},\ \Eprint {http://arxiv.org/abs/1602.03837} {arXiv:1602.03837 [gr-qc]} \BibitemShut {NoStop}%
\bibitem [{\citenamefont {Abbott}\ \emph {et~al.}(2019)\citenamefont {Abbott} \emph {et~al.}}]{LIGOScientific:2018jsj}%
  \BibitemOpen
  \bibfield  {author} {\bibinfo {author} {\bibfnamefont {B.~P.}\ \bibnamefont {Abbott}} \emph {et~al.} (\bibinfo {collaboration} {LIGO Scientific, Virgo}),\ }\href {\doibase 10.3847/2041-8213/ab3800} {\bibfield  {journal} {\bibinfo  {journal} {Astrophys. J. Lett.}\ }\textbf {\bibinfo {volume} {882}},\ \bibinfo {pages} {L24} (\bibinfo {year} {2019})},\ \Eprint {http://arxiv.org/abs/1811.12940} {arXiv:1811.12940 [astro-ph.HE]} \BibitemShut {NoStop}%
\bibitem [{\citenamefont {Akiyama}\ \emph {et~al.}(2019)\citenamefont {Akiyama} \emph {et~al.}}]{EventHorizonTelescope:2019dse}%
  \BibitemOpen
  \bibfield  {author} {\bibinfo {author} {\bibfnamefont {K.}~\bibnamefont {Akiyama}} \emph {et~al.} (\bibinfo {collaboration} {Event Horizon Telescope}),\ }\href {\doibase 10.3847/2041-8213/ab0ec7} {\bibfield  {journal} {\bibinfo  {journal} {Astrophys. J. Lett.}\ }\textbf {\bibinfo {volume} {875}},\ \bibinfo {pages} {L1} (\bibinfo {year} {2019})},\ \Eprint {http://arxiv.org/abs/1906.11238} {arXiv:1906.11238 [astro-ph.GA]} \BibitemShut {NoStop}%
\bibitem [{\citenamefont {Akiyama}\ \emph {et~al.}(2022)\citenamefont {Akiyama} \emph {et~al.}}]{EventHorizonTelescope:2022wkp}%
  \BibitemOpen
  \bibfield  {author} {\bibinfo {author} {\bibfnamefont {K.}~\bibnamefont {Akiyama}} \emph {et~al.} (\bibinfo {collaboration} {Event Horizon Telescope}),\ }\href {\doibase 10.3847/2041-8213/ac6674} {\bibfield  {journal} {\bibinfo  {journal} {Astrophys. J. Lett.}\ }\textbf {\bibinfo {volume} {930}},\ \bibinfo {pages} {L12} (\bibinfo {year} {2022})},\ \Eprint {http://arxiv.org/abs/2311.08680} {arXiv:2311.08680 [astro-ph.HE]} \BibitemShut {NoStop}%
\bibitem [{\citenamefont {Israel}(1967)}]{Israel:1967wq}%
  \BibitemOpen
  \bibfield  {author} {\bibinfo {author} {\bibfnamefont {W.}~\bibnamefont {Israel}},\ }\href {\doibase 10.1103/PhysRev.164.1776} {\bibfield  {journal} {\bibinfo  {journal} {Phys. Rev.}\ }\textbf {\bibinfo {volume} {164}},\ \bibinfo {pages} {1776} (\bibinfo {year} {1967})}\BibitemShut {NoStop}%
\bibitem [{\citenamefont {Carter}(1971)}]{Carter:1971zc}%
  \BibitemOpen
  \bibfield  {author} {\bibinfo {author} {\bibfnamefont {B.}~\bibnamefont {Carter}},\ }\href {\doibase 10.1103/PhysRevLett.26.331} {\bibfield  {journal} {\bibinfo  {journal} {Phys. Rev. Lett.}\ }\textbf {\bibinfo {volume} {26}},\ \bibinfo {pages} {331} (\bibinfo {year} {1971})}\BibitemShut {NoStop}%
\bibitem [{\citenamefont {Robinson}(1975)}]{Robinson:1975bv}%
  \BibitemOpen
  \bibfield  {author} {\bibinfo {author} {\bibfnamefont {D.~C.}\ \bibnamefont {Robinson}},\ }\href {\doibase 10.1103/PhysRevLett.34.905} {\bibfield  {journal} {\bibinfo  {journal} {Phys. Rev. Lett.}\ }\textbf {\bibinfo {volume} {34}},\ \bibinfo {pages} {905} (\bibinfo {year} {1975})}\BibitemShut {NoStop}%
\bibitem [{\citenamefont {Bekenstein}(1995)}]{Bekenstein:1995un}%
  \BibitemOpen
  \bibfield  {author} {\bibinfo {author} {\bibfnamefont {J.~D.}\ \bibnamefont {Bekenstein}},\ }\href {\doibase 10.1103/PhysRevD.51.R6608} {\bibfield  {journal} {\bibinfo  {journal} {Phys. Rev. D}\ }\textbf {\bibinfo {volume} {51}},\ \bibinfo {pages} {R6608} (\bibinfo {year} {1995})}\BibitemShut {NoStop}%
\bibitem [{\citenamefont {Sotiriou}(2015)}]{Sotiriou:2015pka}%
  \BibitemOpen
  \bibfield  {author} {\bibinfo {author} {\bibfnamefont {T.~P.}\ \bibnamefont {Sotiriou}},\ }\href {\doibase 10.1088/0264-9381/32/21/214002} {\bibfield  {journal} {\bibinfo  {journal} {Class. Quant. Grav.}\ }\textbf {\bibinfo {volume} {32}},\ \bibinfo {pages} {214002} (\bibinfo {year} {2015})},\ \Eprint {http://arxiv.org/abs/1505.00248} {arXiv:1505.00248 [gr-qc]} \BibitemShut {NoStop}%
\bibitem [{\citenamefont {Herdeiro}\ and\ \citenamefont {Radu}(2015)}]{Herdeiro:2015waa}%
  \BibitemOpen
  \bibfield  {author} {\bibinfo {author} {\bibfnamefont {C.~A.~R.}\ \bibnamefont {Herdeiro}}\ and\ \bibinfo {author} {\bibfnamefont {E.}~\bibnamefont {Radu}},\ }\href {\doibase 10.1142/S0218271815420146} {\bibfield  {journal} {\bibinfo  {journal} {Int. J. Mod. Phys. D}\ }\textbf {\bibinfo {volume} {24}},\ \bibinfo {pages} {1542014} (\bibinfo {year} {2015})},\ \Eprint {http://arxiv.org/abs/1504.08209} {arXiv:1504.08209 [gr-qc]} \BibitemShut {NoStop}%
\bibitem [{\citenamefont {Kanti}\ \emph {et~al.}(1996)\citenamefont {Kanti}, \citenamefont {Mavromatos}, \citenamefont {Rizos}, \citenamefont {Tamvakis},\ and\ \citenamefont {Winstanley}}]{Kanti:1995vq}%
  \BibitemOpen
  \bibfield  {author} {\bibinfo {author} {\bibfnamefont {P.}~\bibnamefont {Kanti}}, \bibinfo {author} {\bibfnamefont {N.~E.}\ \bibnamefont {Mavromatos}}, \bibinfo {author} {\bibfnamefont {J.}~\bibnamefont {Rizos}}, \bibinfo {author} {\bibfnamefont {K.}~\bibnamefont {Tamvakis}}, \ and\ \bibinfo {author} {\bibfnamefont {E.}~\bibnamefont {Winstanley}},\ }\href {\doibase 10.1103/PhysRevD.54.5049} {\bibfield  {journal} {\bibinfo  {journal} {Phys. Rev. D}\ }\textbf {\bibinfo {volume} {54}},\ \bibinfo {pages} {5049} (\bibinfo {year} {1996})},\ \Eprint {http://arxiv.org/abs/hep-th/9511071} {arXiv:hep-th/9511071} \BibitemShut {NoStop}%
\bibitem [{\citenamefont {Sotiriou}\ and\ \citenamefont {Zhou}(2014)}]{Sotiriou:2013qea}%
  \BibitemOpen
  \bibfield  {author} {\bibinfo {author} {\bibfnamefont {T.~P.}\ \bibnamefont {Sotiriou}}\ and\ \bibinfo {author} {\bibfnamefont {S.-Y.}\ \bibnamefont {Zhou}},\ }\href {\doibase 10.1103/PhysRevLett.112.251102} {\bibfield  {journal} {\bibinfo  {journal} {Phys. Rev. Lett.}\ }\textbf {\bibinfo {volume} {112}},\ \bibinfo {pages} {251102} (\bibinfo {year} {2014})},\ \Eprint {http://arxiv.org/abs/1312.3622} {arXiv:1312.3622 [gr-qc]} \BibitemShut {NoStop}%
\bibitem [{\citenamefont {Antoniou}\ \emph {et~al.}(2018)\citenamefont {Antoniou}, \citenamefont {Bakopoulos},\ and\ \citenamefont {Kanti}}]{Antoniou:2017acq}%
  \BibitemOpen
  \bibfield  {author} {\bibinfo {author} {\bibfnamefont {G.}~\bibnamefont {Antoniou}}, \bibinfo {author} {\bibfnamefont {A.}~\bibnamefont {Bakopoulos}}, \ and\ \bibinfo {author} {\bibfnamefont {P.}~\bibnamefont {Kanti}},\ }\href {\doibase 10.1103/PhysRevLett.120.131102} {\bibfield  {journal} {\bibinfo  {journal} {Phys. Rev. Lett.}\ }\textbf {\bibinfo {volume} {120}},\ \bibinfo {pages} {131102} (\bibinfo {year} {2018})},\ \Eprint {http://arxiv.org/abs/1711.03390} {arXiv:1711.03390 [hep-th]} \BibitemShut {NoStop}%
\bibitem [{\citenamefont {Doneva}\ and\ \citenamefont {Yazadjiev}(2018)}]{Doneva:2017bvd}%
  \BibitemOpen
  \bibfield  {author} {\bibinfo {author} {\bibfnamefont {D.~D.}\ \bibnamefont {Doneva}}\ and\ \bibinfo {author} {\bibfnamefont {S.~S.}\ \bibnamefont {Yazadjiev}},\ }\href {\doibase 10.1103/PhysRevLett.120.131103} {\bibfield  {journal} {\bibinfo  {journal} {Phys. Rev. Lett.}\ }\textbf {\bibinfo {volume} {120}},\ \bibinfo {pages} {131103} (\bibinfo {year} {2018})},\ \Eprint {http://arxiv.org/abs/1711.01187} {arXiv:1711.01187 [gr-qc]} \BibitemShut {NoStop}%
\bibitem [{\citenamefont {Silva}\ \emph {et~al.}(2018)\citenamefont {Silva}, \citenamefont {Sakstein}, \citenamefont {Gualtieri}, \citenamefont {Sotiriou},\ and\ \citenamefont {Berti}}]{Silva:2017uqg}%
  \BibitemOpen
  \bibfield  {author} {\bibinfo {author} {\bibfnamefont {H.~O.}\ \bibnamefont {Silva}}, \bibinfo {author} {\bibfnamefont {J.}~\bibnamefont {Sakstein}}, \bibinfo {author} {\bibfnamefont {L.}~\bibnamefont {Gualtieri}}, \bibinfo {author} {\bibfnamefont {T.~P.}\ \bibnamefont {Sotiriou}}, \ and\ \bibinfo {author} {\bibfnamefont {E.}~\bibnamefont {Berti}},\ }\href {\doibase 10.1103/PhysRevLett.120.131104} {\bibfield  {journal} {\bibinfo  {journal} {Phys. Rev. Lett.}\ }\textbf {\bibinfo {volume} {120}},\ \bibinfo {pages} {131104} (\bibinfo {year} {2018})},\ \Eprint {http://arxiv.org/abs/1711.02080} {arXiv:1711.02080 [gr-qc]} \BibitemShut {NoStop}%
\bibitem [{\citenamefont {Cunha}\ \emph {et~al.}(2019)\citenamefont {Cunha}, \citenamefont {Herdeiro},\ and\ \citenamefont {Radu}}]{Cunha:2019dwb}%
  \BibitemOpen
  \bibfield  {author} {\bibinfo {author} {\bibfnamefont {P.~V.~P.}\ \bibnamefont {Cunha}}, \bibinfo {author} {\bibfnamefont {C.~A.~R.}\ \bibnamefont {Herdeiro}}, \ and\ \bibinfo {author} {\bibfnamefont {E.}~\bibnamefont {Radu}},\ }\href {\doibase 10.1103/PhysRevLett.123.011101} {\bibfield  {journal} {\bibinfo  {journal} {Phys. Rev. Lett.}\ }\textbf {\bibinfo {volume} {123}},\ \bibinfo {pages} {011101} (\bibinfo {year} {2019})},\ \Eprint {http://arxiv.org/abs/1904.09997} {arXiv:1904.09997 [gr-qc]} \BibitemShut {NoStop}%
\bibitem [{\citenamefont {Collodel}\ \emph {et~al.}(2020)\citenamefont {Collodel}, \citenamefont {Kleihaus}, \citenamefont {Kunz},\ and\ \citenamefont {Berti}}]{Collodel:2019kkx}%
  \BibitemOpen
  \bibfield  {author} {\bibinfo {author} {\bibfnamefont {L.~G.}\ \bibnamefont {Collodel}}, \bibinfo {author} {\bibfnamefont {B.}~\bibnamefont {Kleihaus}}, \bibinfo {author} {\bibfnamefont {J.}~\bibnamefont {Kunz}}, \ and\ \bibinfo {author} {\bibfnamefont {E.}~\bibnamefont {Berti}},\ }\href {\doibase 10.1088/1361-6382/ab74f9} {\bibfield  {journal} {\bibinfo  {journal} {Class. Quant. Grav.}\ }\textbf {\bibinfo {volume} {37}},\ \bibinfo {pages} {075018} (\bibinfo {year} {2020})},\ \Eprint {http://arxiv.org/abs/1912.05382} {arXiv:1912.05382 [gr-qc]} \BibitemShut {NoStop}%
\bibitem [{\citenamefont {Dima}\ \emph {et~al.}(2020)\citenamefont {Dima}, \citenamefont {Barausse}, \citenamefont {Franchini},\ and\ \citenamefont {Sotiriou}}]{Dima:2020yac}%
  \BibitemOpen
  \bibfield  {author} {\bibinfo {author} {\bibfnamefont {A.}~\bibnamefont {Dima}}, \bibinfo {author} {\bibfnamefont {E.}~\bibnamefont {Barausse}}, \bibinfo {author} {\bibfnamefont {N.}~\bibnamefont {Franchini}}, \ and\ \bibinfo {author} {\bibfnamefont {T.~P.}\ \bibnamefont {Sotiriou}},\ }\href {\doibase 10.1103/PhysRevLett.125.231101} {\bibfield  {journal} {\bibinfo  {journal} {Phys. Rev. Lett.}\ }\textbf {\bibinfo {volume} {125}},\ \bibinfo {pages} {231101} (\bibinfo {year} {2020})},\ \Eprint {http://arxiv.org/abs/2006.03095} {arXiv:2006.03095 [gr-qc]} \BibitemShut {NoStop}%
\bibitem [{\citenamefont {Herdeiro}\ \emph {et~al.}(2021)\citenamefont {Herdeiro}, \citenamefont {Radu}, \citenamefont {Silva}, \citenamefont {Sotiriou},\ and\ \citenamefont {Yunes}}]{Herdeiro:2020wei}%
  \BibitemOpen
  \bibfield  {author} {\bibinfo {author} {\bibfnamefont {C.~A.~R.}\ \bibnamefont {Herdeiro}}, \bibinfo {author} {\bibfnamefont {E.}~\bibnamefont {Radu}}, \bibinfo {author} {\bibfnamefont {H.~O.}\ \bibnamefont {Silva}}, \bibinfo {author} {\bibfnamefont {T.~P.}\ \bibnamefont {Sotiriou}}, \ and\ \bibinfo {author} {\bibfnamefont {N.}~\bibnamefont {Yunes}},\ }\href {\doibase 10.1103/PhysRevLett.126.011103} {\bibfield  {journal} {\bibinfo  {journal} {Phys. Rev. Lett.}\ }\textbf {\bibinfo {volume} {126}},\ \bibinfo {pages} {011103} (\bibinfo {year} {2021})},\ \Eprint {http://arxiv.org/abs/2009.03904} {arXiv:2009.03904 [gr-qc]} \BibitemShut {NoStop}%
\bibitem [{\citenamefont {Berti}\ \emph {et~al.}(2021)\citenamefont {Berti}, \citenamefont {Collodel}, \citenamefont {Kleihaus},\ and\ \citenamefont {Kunz}}]{Berti:2020kgk}%
  \BibitemOpen
  \bibfield  {author} {\bibinfo {author} {\bibfnamefont {E.}~\bibnamefont {Berti}}, \bibinfo {author} {\bibfnamefont {L.~G.}\ \bibnamefont {Collodel}}, \bibinfo {author} {\bibfnamefont {B.}~\bibnamefont {Kleihaus}}, \ and\ \bibinfo {author} {\bibfnamefont {J.}~\bibnamefont {Kunz}},\ }\href {\doibase 10.1103/PhysRevLett.126.011104} {\bibfield  {journal} {\bibinfo  {journal} {Phys. Rev. Lett.}\ }\textbf {\bibinfo {volume} {126}},\ \bibinfo {pages} {011104} (\bibinfo {year} {2021})},\ \Eprint {http://arxiv.org/abs/2009.03905} {arXiv:2009.03905 [gr-qc]} \BibitemShut {NoStop}%
\bibitem [{\citenamefont {Herdeiro}\ \emph {et~al.}(2018)\citenamefont {Herdeiro}, \citenamefont {Radu}, \citenamefont {Sanchis-Gual},\ and\ \citenamefont {Font}}]{Herdeiro:2018wub}%
  \BibitemOpen
  \bibfield  {author} {\bibinfo {author} {\bibfnamefont {C.~A.~R.}\ \bibnamefont {Herdeiro}}, \bibinfo {author} {\bibfnamefont {E.}~\bibnamefont {Radu}}, \bibinfo {author} {\bibfnamefont {N.}~\bibnamefont {Sanchis-Gual}}, \ and\ \bibinfo {author} {\bibfnamefont {J.~A.}\ \bibnamefont {Font}},\ }\href {\doibase 10.1103/PhysRevLett.121.101102} {\bibfield  {journal} {\bibinfo  {journal} {Phys. Rev. Lett.}\ }\textbf {\bibinfo {volume} {121}},\ \bibinfo {pages} {101102} (\bibinfo {year} {2018})},\ \Eprint {http://arxiv.org/abs/1806.05190} {arXiv:1806.05190 [gr-qc]} \BibitemShut {NoStop}%
\bibitem [{\citenamefont {Myung}\ and\ \citenamefont {Zou}(2019{\natexlab{a}})}]{Myung:2018vug}%
  \BibitemOpen
  \bibfield  {author} {\bibinfo {author} {\bibfnamefont {Y.~S.}\ \bibnamefont {Myung}}\ and\ \bibinfo {author} {\bibfnamefont {D.-C.}\ \bibnamefont {Zou}},\ }\href {\doibase 10.1140/epjc/s10052-019-6792-6} {\bibfield  {journal} {\bibinfo  {journal} {Eur. Phys. J. C}\ }\textbf {\bibinfo {volume} {79}},\ \bibinfo {pages} {273} (\bibinfo {year} {2019}{\natexlab{a}})},\ \Eprint {http://arxiv.org/abs/1808.02609} {arXiv:1808.02609 [gr-qc]} \BibitemShut {NoStop}%
\bibitem [{\citenamefont {Myung}\ and\ \citenamefont {Zou}(2019{\natexlab{b}})}]{Myung:2018jvi}%
  \BibitemOpen
  \bibfield  {author} {\bibinfo {author} {\bibfnamefont {Y.~S.}\ \bibnamefont {Myung}}\ and\ \bibinfo {author} {\bibfnamefont {D.-C.}\ \bibnamefont {Zou}},\ }\href {\doibase 10.1016/j.physletb.2019.01.046} {\bibfield  {journal} {\bibinfo  {journal} {Phys. Lett. B}\ }\textbf {\bibinfo {volume} {790}},\ \bibinfo {pages} {400} (\bibinfo {year} {2019}{\natexlab{b}})},\ \Eprint {http://arxiv.org/abs/1812.03604} {arXiv:1812.03604 [gr-qc]} \BibitemShut {NoStop}%
\bibitem [{\citenamefont {Fernandes}\ \emph {et~al.}(2019)\citenamefont {Fernandes}, \citenamefont {Herdeiro}, \citenamefont {Pombo}, \citenamefont {Radu},\ and\ \citenamefont {Sanchis-Gual}}]{Fernandes:2019rez}%
  \BibitemOpen
  \bibfield  {author} {\bibinfo {author} {\bibfnamefont {P.~G.~S.}\ \bibnamefont {Fernandes}}, \bibinfo {author} {\bibfnamefont {C.~A.~R.}\ \bibnamefont {Herdeiro}}, \bibinfo {author} {\bibfnamefont {A.~M.}\ \bibnamefont {Pombo}}, \bibinfo {author} {\bibfnamefont {E.}~\bibnamefont {Radu}}, \ and\ \bibinfo {author} {\bibfnamefont {N.}~\bibnamefont {Sanchis-Gual}},\ }\href {\doibase 10.1088/1361-6382/ab23a1} {\bibfield  {journal} {\bibinfo  {journal} {Class. Quant. Grav.}\ }\textbf {\bibinfo {volume} {36}},\ \bibinfo {pages} {134002} (\bibinfo {year} {2019})},\ \bibinfo {note} {[Erratum: Class.Quant.Grav. 37, 049501 (2020)]},\ \Eprint {http://arxiv.org/abs/1902.05079} {arXiv:1902.05079 [gr-qc]} \BibitemShut {NoStop}%
\bibitem [{\citenamefont {Astefanesei}\ \emph {et~al.}(2019)\citenamefont {Astefanesei}, \citenamefont {Herdeiro}, \citenamefont {Pombo},\ and\ \citenamefont {Radu}}]{Astefanesei:2019pfq}%
  \BibitemOpen
  \bibfield  {author} {\bibinfo {author} {\bibfnamefont {D.}~\bibnamefont {Astefanesei}}, \bibinfo {author} {\bibfnamefont {C.}~\bibnamefont {Herdeiro}}, \bibinfo {author} {\bibfnamefont {A.}~\bibnamefont {Pombo}}, \ and\ \bibinfo {author} {\bibfnamefont {E.}~\bibnamefont {Radu}},\ }\href {\doibase 10.1007/JHEP10(2019)078} {\bibfield  {journal} {\bibinfo  {journal} {JHEP}\ }\textbf {\bibinfo {volume} {10}},\ \bibinfo {pages} {078} (\bibinfo {year} {2019})},\ \Eprint {http://arxiv.org/abs/1905.08304} {arXiv:1905.08304 [hep-th]} \BibitemShut {NoStop}%
\bibitem [{\citenamefont {Bl{\'a}zquez-Salcedo}\ \emph {et~al.}(2020)\citenamefont {Bl{\'a}zquez-Salcedo}, \citenamefont {Herdeiro}, \citenamefont {Kunz}, \citenamefont {Pombo},\ and\ \citenamefont {Radu}}]{Blazquez-Salcedo:2020nhs}%
  \BibitemOpen
  \bibfield  {author} {\bibinfo {author} {\bibfnamefont {J.~L.}\ \bibnamefont {Bl{\'a}zquez-Salcedo}}, \bibinfo {author} {\bibfnamefont {C.~A.~R.}\ \bibnamefont {Herdeiro}}, \bibinfo {author} {\bibfnamefont {J.}~\bibnamefont {Kunz}}, \bibinfo {author} {\bibfnamefont {A.~M.}\ \bibnamefont {Pombo}}, \ and\ \bibinfo {author} {\bibfnamefont {E.}~\bibnamefont {Radu}},\ }\href {\doibase 10.1016/j.physletb.2020.135493} {\bibfield  {journal} {\bibinfo  {journal} {Phys. Lett. B}\ }\textbf {\bibinfo {volume} {806}},\ \bibinfo {pages} {135493} (\bibinfo {year} {2020})},\ \Eprint {http://arxiv.org/abs/2002.00963} {arXiv:2002.00963 [gr-qc]} \BibitemShut {NoStop}%
\bibitem [{\citenamefont {Chen}\ \emph {et~al.}(2026)\citenamefont {Chen}, \citenamefont {Chew},\ and\ \citenamefont {Kunz}}]{Chen:2026olq}%
  \BibitemOpen
  \bibfield  {author} {\bibinfo {author} {\bibfnamefont {S.}~\bibnamefont {Chen}}, \bibinfo {author} {\bibfnamefont {X.~Y.}\ \bibnamefont {Chew}}, \ and\ \bibinfo {author} {\bibfnamefont {J.}~\bibnamefont {Kunz}},\ }\href@noop {} {\  (\bibinfo {year} {2026})},\ \Eprint {http://arxiv.org/abs/2603.16701} {arXiv:2603.16701 [gr-qc]} \BibitemShut {NoStop}%
\bibitem [{\citenamefont {Doneva}\ and\ \citenamefont {Yazadjiev}(2021)}]{Doneva:2021dcc}%
  \BibitemOpen
  \bibfield  {author} {\bibinfo {author} {\bibfnamefont {D.~D.}\ \bibnamefont {Doneva}}\ and\ \bibinfo {author} {\bibfnamefont {S.~S.}\ \bibnamefont {Yazadjiev}},\ }\href {\doibase 10.1103/PhysRevD.103.083007} {\bibfield  {journal} {\bibinfo  {journal} {Phys. Rev. D}\ }\textbf {\bibinfo {volume} {103}},\ \bibinfo {pages} {083007} (\bibinfo {year} {2021})},\ \Eprint {http://arxiv.org/abs/2102.03940} {arXiv:2102.03940 [gr-qc]} \BibitemShut {NoStop}%
\bibitem [{\citenamefont {Fan}\ \emph {et~al.}(2024)\citenamefont {Fan}, \citenamefont {Myung}, \citenamefont {Zou},\ and\ \citenamefont {Lai}}]{Fan:2023jhi}%
  \BibitemOpen
  \bibfield  {author} {\bibinfo {author} {\bibfnamefont {K.-H.}\ \bibnamefont {Fan}}, \bibinfo {author} {\bibfnamefont {Y.~S.}\ \bibnamefont {Myung}}, \bibinfo {author} {\bibfnamefont {D.-C.}\ \bibnamefont {Zou}}, \ and\ \bibinfo {author} {\bibfnamefont {M.-Y.}\ \bibnamefont {Lai}},\ }\href {\doibase 10.1140/epjc/s10052-024-13030-y} {\bibfield  {journal} {\bibinfo  {journal} {Eur. Phys. J. C}\ }\textbf {\bibinfo {volume} {84}},\ \bibinfo {pages} {679} (\bibinfo {year} {2024})},\ \Eprint {http://arxiv.org/abs/2401.00144} {arXiv:2401.00144 [gr-qc]} \BibitemShut {NoStop}%
\bibitem [{\citenamefont {Liu}\ and\ \citenamefont {Zhang}(2024)}]{Liu:2024bzh}%
  \BibitemOpen
  \bibfield  {author} {\bibinfo {author} {\bibfnamefont {H.-S.}\ \bibnamefont {Liu}}\ and\ \bibinfo {author} {\bibfnamefont {L.}~\bibnamefont {Zhang}},\ }\href {\doibase 10.1007/JHEP10(2024)067} {\bibfield  {journal} {\bibinfo  {journal} {JHEP}\ }\textbf {\bibinfo {volume} {10}},\ \bibinfo {pages} {067} (\bibinfo {year} {2024})},\ \Eprint {http://arxiv.org/abs/2407.08208} {arXiv:2407.08208 [gr-qc]} \BibitemShut {NoStop}%
\bibitem [{\citenamefont {Zhang}\ \emph {et~al.}(2024)\citenamefont {Zhang}, \citenamefont {Pan}, \citenamefont {Myung},\ and\ \citenamefont {Zou}}]{Zhang:2024bfu}%
  \BibitemOpen
  \bibfield  {author} {\bibinfo {author} {\bibfnamefont {L.}~\bibnamefont {Zhang}}, \bibinfo {author} {\bibfnamefont {Q.}~\bibnamefont {Pan}}, \bibinfo {author} {\bibfnamefont {Y.~S.}\ \bibnamefont {Myung}}, \ and\ \bibinfo {author} {\bibfnamefont {D.-C.}\ \bibnamefont {Zou}},\ }\href {\doibase 10.1103/PhysRevD.110.124036} {\bibfield  {journal} {\bibinfo  {journal} {Phys. Rev. D}\ }\textbf {\bibinfo {volume} {110}},\ \bibinfo {pages} {124036} (\bibinfo {year} {2024})},\ \Eprint {http://arxiv.org/abs/2409.11669} {arXiv:2409.11669 [gr-qc]} \BibitemShut {NoStop}%
\bibitem [{\citenamefont {Hod}(2020)}]{Hod:2020jjy}%
  \BibitemOpen
  \bibfield  {author} {\bibinfo {author} {\bibfnamefont {S.}~\bibnamefont {Hod}},\ }\href {\doibase 10.1103/PhysRevD.102.084060} {\bibfield  {journal} {\bibinfo  {journal} {Phys. Rev. D}\ }\textbf {\bibinfo {volume} {102}},\ \bibinfo {pages} {084060} (\bibinfo {year} {2020})},\ \Eprint {http://arxiv.org/abs/2006.09399} {arXiv:2006.09399 [gr-qc]} \BibitemShut {NoStop}%
\bibitem [{\citenamefont {Doneva}\ and\ \citenamefont {Yazadjiev}(2022)}]{Doneva:2021tvn}%
  \BibitemOpen
  \bibfield  {author} {\bibinfo {author} {\bibfnamefont {D.~D.}\ \bibnamefont {Doneva}}\ and\ \bibinfo {author} {\bibfnamefont {S.~S.}\ \bibnamefont {Yazadjiev}},\ }\href {\doibase 10.1103/PhysRevD.105.L041502} {\bibfield  {journal} {\bibinfo  {journal} {Phys. Rev. D}\ }\textbf {\bibinfo {volume} {105}},\ \bibinfo {pages} {L041502} (\bibinfo {year} {2022})},\ \Eprint {http://arxiv.org/abs/2107.01738} {arXiv:2107.01738 [gr-qc]} \BibitemShut {NoStop}%
\bibitem [{\citenamefont {Doneva}\ \emph {et~al.}(2022)\citenamefont {Doneva}, \citenamefont {Collodel},\ and\ \citenamefont {Yazadjiev}}]{Doneva:2022yqu}%
  \BibitemOpen
  \bibfield  {author} {\bibinfo {author} {\bibfnamefont {D.~D.}\ \bibnamefont {Doneva}}, \bibinfo {author} {\bibfnamefont {L.~G.}\ \bibnamefont {Collodel}}, \ and\ \bibinfo {author} {\bibfnamefont {S.~S.}\ \bibnamefont {Yazadjiev}},\ }\href {\doibase 10.1103/PhysRevD.106.104027} {\bibfield  {journal} {\bibinfo  {journal} {Phys. Rev. D}\ }\textbf {\bibinfo {volume} {106}},\ \bibinfo {pages} {104027} (\bibinfo {year} {2022})},\ \Eprint {http://arxiv.org/abs/2208.02077} {arXiv:2208.02077 [gr-qc]} \BibitemShut {NoStop}%
\bibitem [{\citenamefont {Lai}\ \emph {et~al.}(2023)\citenamefont {Lai}, \citenamefont {Zou}, \citenamefont {Yue},\ and\ \citenamefont {Myung}}]{Lai:2023gwe}%
  \BibitemOpen
  \bibfield  {author} {\bibinfo {author} {\bibfnamefont {M.-Y.}\ \bibnamefont {Lai}}, \bibinfo {author} {\bibfnamefont {D.-C.}\ \bibnamefont {Zou}}, \bibinfo {author} {\bibfnamefont {R.-H.}\ \bibnamefont {Yue}}, \ and\ \bibinfo {author} {\bibfnamefont {Y.~S.}\ \bibnamefont {Myung}},\ }\href {\doibase 10.1103/PhysRevD.108.084007} {\bibfield  {journal} {\bibinfo  {journal} {Phys. Rev. D}\ }\textbf {\bibinfo {volume} {108}},\ \bibinfo {pages} {084007} (\bibinfo {year} {2023})},\ \Eprint {http://arxiv.org/abs/2304.08012} {arXiv:2304.08012 [gr-qc]} \BibitemShut {NoStop}%
\bibitem [{\citenamefont {Zou}\ \emph {et~al.}(2024)\citenamefont {Zou}, \citenamefont {Yang}, \citenamefont {Lai}, \citenamefont {Huang}, \citenamefont {Liu}, \citenamefont {Kunz}, \citenamefont {Myung},\ and\ \citenamefont {Yue}}]{Zou:2024wsk}%
  \BibitemOpen
  \bibfield  {author} {\bibinfo {author} {\bibfnamefont {D.-C.}\ \bibnamefont {Zou}}, \bibinfo {author} {\bibfnamefont {X.}~\bibnamefont {Yang}}, \bibinfo {author} {\bibfnamefont {M.-Y.}\ \bibnamefont {Lai}}, \bibinfo {author} {\bibfnamefont {H.}~\bibnamefont {Huang}}, \bibinfo {author} {\bibfnamefont {B.}~\bibnamefont {Liu}}, \bibinfo {author} {\bibfnamefont {J.}~\bibnamefont {Kunz}}, \bibinfo {author} {\bibfnamefont {Y.~S.}\ \bibnamefont {Myung}}, \ and\ \bibinfo {author} {\bibfnamefont {R.-H.}\ \bibnamefont {Yue}},\ }\href@noop {} {\  (\bibinfo {year} {2024})},\ \Eprint {http://arxiv.org/abs/2404.19521} {arXiv:2404.19521 [gr-qc]} \BibitemShut {NoStop}%
\bibitem [{\citenamefont {Liu}\ \emph {et~al.}(2025)\citenamefont {Liu}, \citenamefont {Liu}, \citenamefont {Peng},\ and\ \citenamefont {Zhang}}]{Liu:2025eve}%
  \BibitemOpen
  \bibfield  {author} {\bibinfo {author} {\bibfnamefont {S.}~\bibnamefont {Liu}}, \bibinfo {author} {\bibfnamefont {Y.}~\bibnamefont {Liu}}, \bibinfo {author} {\bibfnamefont {Y.}~\bibnamefont {Peng}}, \ and\ \bibinfo {author} {\bibfnamefont {C.-Y.}\ \bibnamefont {Zhang}},\ }\href {\doibase 10.1140/epjc/s10052-025-15096-8} {\bibfield  {journal} {\bibinfo  {journal} {Eur. Phys. J. C}\ }\textbf {\bibinfo {volume} {85}},\ \bibinfo {pages} {1370} (\bibinfo {year} {2025})},\ \Eprint {http://arxiv.org/abs/2509.17892} {arXiv:2509.17892 [gr-qc]} \BibitemShut {NoStop}%
\bibitem [{\citenamefont {Kunz}\ \emph {et~al.}(2019)\citenamefont {Kunz}, \citenamefont {Perapechka},\ and\ \citenamefont {Shnir}}]{Kunz:2019bhm}%
  \BibitemOpen
  \bibfield  {author} {\bibinfo {author} {\bibfnamefont {J.}~\bibnamefont {Kunz}}, \bibinfo {author} {\bibfnamefont {I.}~\bibnamefont {Perapechka}}, \ and\ \bibinfo {author} {\bibfnamefont {Y.}~\bibnamefont {Shnir}},\ }\href {\doibase 10.1103/PhysRevD.100.064032} {\bibfield  {journal} {\bibinfo  {journal} {Phys. Rev. D}\ }\textbf {\bibinfo {volume} {100}},\ \bibinfo {pages} {064032} (\bibinfo {year} {2019})},\ \Eprint {http://arxiv.org/abs/1904.07630} {arXiv:1904.07630 [gr-qc]} \BibitemShut {NoStop}%
\bibitem [{\citenamefont {Fernandes}\ and\ \citenamefont {Mulryne}(2023)}]{Fernandes:2022gde}%
  \BibitemOpen
  \bibfield  {author} {\bibinfo {author} {\bibfnamefont {P.~G.~S.}\ \bibnamefont {Fernandes}}\ and\ \bibinfo {author} {\bibfnamefont {D.~J.}\ \bibnamefont {Mulryne}},\ }\href {\doibase 10.1088/1361-6382/ace232} {\bibfield  {journal} {\bibinfo  {journal} {Class. Quant. Grav.}\ }\textbf {\bibinfo {volume} {40}},\ \bibinfo {pages} {165001} (\bibinfo {year} {2023})},\ \Eprint {http://arxiv.org/abs/2212.07293} {arXiv:2212.07293 [gr-qc]} \BibitemShut {NoStop}%
\bibitem [{\citenamefont {Silva}\ \emph {et~al.}(2019)\citenamefont {Silva}, \citenamefont {Macedo}, \citenamefont {Sotiriou}, \citenamefont {Gualtieri}, \citenamefont {Sakstein},\ and\ \citenamefont {Berti}}]{Silva:2018qhn}%
  \BibitemOpen
  \bibfield  {author} {\bibinfo {author} {\bibfnamefont {H.~O.}\ \bibnamefont {Silva}}, \bibinfo {author} {\bibfnamefont {C.~F.~B.}\ \bibnamefont {Macedo}}, \bibinfo {author} {\bibfnamefont {T.~P.}\ \bibnamefont {Sotiriou}}, \bibinfo {author} {\bibfnamefont {L.}~\bibnamefont {Gualtieri}}, \bibinfo {author} {\bibfnamefont {J.}~\bibnamefont {Sakstein}}, \ and\ \bibinfo {author} {\bibfnamefont {E.}~\bibnamefont {Berti}},\ }\href {\doibase 10.1103/PhysRevD.99.064011} {\bibfield  {journal} {\bibinfo  {journal} {Phys. Rev. D}\ }\textbf {\bibinfo {volume} {99}},\ \bibinfo {pages} {064011} (\bibinfo {year} {2019})},\ \Eprint {http://arxiv.org/abs/1812.05590} {arXiv:1812.05590 [gr-qc]} \BibitemShut {NoStop}%
\bibitem [{\citenamefont {Herdeiro}\ \emph {et~al.}(2026)\citenamefont {Herdeiro}, \citenamefont {Huang}, \citenamefont {Kunz}, \citenamefont {Lai}, \citenamefont {Radu},\ and\ \citenamefont {Zou}}]{Herdeiro:2026sur}%
  \BibitemOpen
  \bibfield  {author} {\bibinfo {author} {\bibfnamefont {C.}~\bibnamefont {Herdeiro}}, \bibinfo {author} {\bibfnamefont {H.}~\bibnamefont {Huang}}, \bibinfo {author} {\bibfnamefont {J.}~\bibnamefont {Kunz}}, \bibinfo {author} {\bibfnamefont {M.-Y.}\ \bibnamefont {Lai}}, \bibinfo {author} {\bibfnamefont {E.}~\bibnamefont {Radu}}, \ and\ \bibinfo {author} {\bibfnamefont {D.-C.}\ \bibnamefont {Zou}},\ }\href@noop {} {\  (\bibinfo {year} {2026})},\ \Eprint {http://arxiv.org/abs/2603.24164} {arXiv:2603.24164 [gr-qc]} \BibitemShut {NoStop}%
\end{thebibliography}%

\end{document}